\documentstyle[12pt,epsf]{article}
\newcommand{\be}{\begin{equation}}
\newcommand{\en}{\end{equation}}
\newcommand{\bea}{\begin{eqnarray}}
\newcommand{\ena}{\end{eqnarray}}
\newcommand{\Det}{\hbox{Det}}
\newcommand{\dete}{\hbox{det}}
\newcommand{\hbo}{\hbox to 1 true cm {\hfill } }
\newcommand{\tr}{\hbox{tr}}

\def\dslash{\partial\kern-.6em\slash}
\def\kslash{k\kern-.5em\slash}
\def\pslash{p\kern-.4em\slash}
\def\Dslash{D\kern-.6em\slash}
\def\Vslash{V\kern-.7em\slash}
\def\vslash{v\kern-.5em\slash}
\def\rslash{r\kern-.5em\slash}
\def\qslash{q\kern-.5em\slash}
\def\Aslash{A\kern-.5em\slash}
\def\pr{Phys. Rev.}
\def\prl{Phys. Rev. Lett.}
\def\np{Nucl. Phys.}

\begin{document}
\vglue 1truecm

\vbox{\hfill KIAS-P98029}

\hfill  May 14, 1999

\vfil
\centerline{\large\bf Quark Condensation, Induced Symmetry
Breaking and }
\centerline{\large\bf Color Superconductivity at High Density }

\bigskip
\centerline{Kurt Langfeld$^{a,b}$ and Mannque Rho$^{a,c}$ }
\vspace{1 true cm}
\centerline{ $^a$  School of Physics, Korea Institute for Advanced
Study}
\centerline{ Seoul 130-012, Korea }
\bigskip
\centerline{ $^b$ Institut f\"ur Theoretische Physik, Universit\"at
   T\"ubingen }
\centerline{D--72076 T\"ubingen, Germany}
\bigskip
\centerline{ $^c$ Service de Physique Th\'eorique, C.E. Saclay, }
\centerline{ F-91191 Gif-sur-Yvette Cedex, France }

\vfil
\begin{abstract}

The phase structure of hadronic matter at high density relevant to
the physics of compact stars and relativistic heavy-ion collisions
is studied in a low-energy effective quark theory. The relevant
phases that figure are (1) chiral condensation, (2) diquark color
condensation (color superconductivity) and (3) induced
Lorentz-symmetry breaking (``ISB").
For a reasonable strength for the {\it effective} four-Fermi
current current interaction implied by the low energy effective quark
theory for systems with a Fermi surface 
we find that the ``ISB" phase sets in together with chiral symmetry
restoration (with the vanishing quark condensate) at a moderate density 
while color superconductivity
associated with scalar diquark condensation is pushed up to an
asymptotic density. Consequently color superconductivity seems rather 
unlikely in heavy-ion collisions although it may play 
a role in compact stars. Lack of confinement in the model makes the
result of this analysis only qualitative but the hierarchy of the
transitions we find seems to be quite robust.

\end{abstract}

\vfil
\hrule width 5truecm
\vskip .2truecm
\begin{quote}
PACS: 11.30.Rd; 12.38.Aw; 

keywords: {\it chiral symmetry breaking; color superconductivity; finite 
density }
\end{quote}
\eject

\section{ Introduction }
\vskip 0.3cm
The dynamics of nucleons in dilute environment at very low energy can be
approached via effective field theories such as chiral perturbation theory.
For instance, the properties of two nucleons, both bound and scattering, can be
very accurately calculated within the framework of well-defined strategy
\cite{pkmr}. The situation with denser systems such as heavy nuclei and
nuclear matter is an entirely different issue. While the traditional
phenomenological methods have
been quite successful in correlating a wide  range of phenomena, there is
very little predictive power in  these techniques in the density regime
where new physics can set in. The question one would like to raise is what
happens to a hadronic (nuclear)
matter system when it is compressed to a high density
as in relativistic heavy ion collisions and
in neutron stars. This requires going, perhaps much, beyond normal nuclear
matter density, $\rho_0\approx 0.16\ {\rm fm}^{-3}$. At some high density,
it is plausible that nuclear matter undergoes one or more phase transitions.
For instance, neutron matter found in neutron stars could be unstable at a
density a few times $\rho_0$ against the condensation of kaons with an
intriguing
consequence on the structure of neutron stars and the formation of black
holes \cite{chlPR,LLB97}. At about the same density or higher, chiral symmetry
restoration with a possible
induced Lorentz symmetry breaking could take place  \cite{lan97,lan98}.

\vskip 0.3cm Even more recently the old idea of color
superconductivity proposed by Bailin and Love \cite{bai84} drew a
revived attention in a modern context and became a focus of wide
speculations and debates \cite{al97,others}.
If this phenomenon were to take place,
it could do so in the regime where either kaon condensation and/or
chiral phase transition and/or induced Lorentz symmetry breaking
might set in and the question is how and in what language,
macroscopic (effective) or microscopic (quark-gluon), adapted to
QCD one should address this problem. So far the most economical
way of describing dense hadronic matter up to and near the normal
nuclear matter density is to treat the hadrons that enter as
quasi-particles, that is, fully dressed particles, with certain
symmetries (such as chiral symmetry and scale invariance) assumed
to be relevant. This is the basis of Brown-Rho scaling
\cite{BR,frs98} which is found to provide one of the simplest
explanations for the low-mass lepton pairs observed in the CERES
experiments \cite{LKB}. The CERES experiments probe densities a
few times the normal matter density, so at least for these data,
the hadronic quasi-particle picture seems to work. However in
dealing with the phase transitions that go over to QCD's
microscopic degrees of freedom, the hadronic quasi-particle
picture should cede to one in which quasi-particles are quarks and
gluons instead. The proper strategy must then be a Fermi-liquid
theory of quarks and the phase transition of the color
superconductivity type could be analyzed in terms of Landau
Fermi-liquid theory and instability from a quark Fermi liquid.

\vskip 0.3cm Such an analysis involves the decimation of energy
scales relative to the Fermi momentum and a thorough study of the
renormalization group flow of the couplings corresponding to
operators which are marginal at the given energy
scale~\cite{pol92}. If these coupling constants grow to order
unity, one would have to resort to intuition to set up the
operators corresponding to the new emerging physics~\cite{pol92}.
In the deconfined phase of QCD at high density, the new physics
which is controlled by one-gluon exchange is most likely the
formation of a color superconducting gap reminiscent of diquark
condensation. For a quantitative investigation of this effect
see~\cite{eva98}. On the other hand, we believe that the
(possible) phase transition due to what we called
``induced Lorentz symmetry breaking (or ISB in short)" 
in refs.\cite{lan97,lan98}
occurs at rather low densities at which QCD is still
strongly coupled. In this case, an educated guess of the relevant
operators as required by the renormalization group analysis is
difficult, and, if the induced symmetry breaking phenomenon
interferes with quark confinement, impossible at the present
stage.

\vskip 0.3cm In this paper we shall therefore adopt a more
coarse-grained point of view and analyze the quark state at finite
density resorting to a low-energy effective quark model. The model
we shall consider consists of a gauge-invariant 
four-quark contact interaction with the quantum numbers of 
one-gluon exchange supplemented in certain channels (e.g., the flavor-singlet
vector channel) with collective modes associated with the presence of a Fermi
sea. Such an interaction represents the lowest-dimension operator that would 
result when the gluon is integrated out from the QCD partition function and 
would be dominant under the renormalization group flow as high energy states,
both elementary and collective, are suitably ``decimated."
 The constraint we shall 
take into account in the current-current interaction -- which controls the
overall strength -- is that it be
consistent with the properties of pseudo-Goldstone bosons (pions)
in matter-free space. The presence of Lorentz non-invariant collective 
excitations could affect significantly the flavor-singlet vector 
channel (that we shall frequently refer to as $V_0$ channel for reasons to be
specified below) relevant to the ISB phase discussed 
in refs.\cite{lan97,lan98}. In
particular, we shall take into account the possibility that the attraction
in this channel could be enhanced by, and intimately correlated with, such
collective modes as $N^* N^{-1}$ excitation considered in \cite{KRBR} for
``dropping" vector masses in dense medium.

\vskip 0.3cm
We admit that the model is plagued with certain shortcomings (e.g.
lack of quark confinement). Nonetheless we believe
it yields a comprehensive view of all
three possible phases, namely, chiral symmetry phase, the ISB
phase and color superconductivity phase.
Given that lattice QCD measurements in the presence of chemical
potential are totally lacking, qualitative studies of the sort we
are presenting here seem to be the best one could do at the
moment.

\vskip 0.3cm
The plan of this paper is as follows. We shall briefly review in
section 2 the scenario of chiral phase restoration at finite
chemical potential with an emphasis of the induced breaking of
Lorentz symmetry developed in ~\cite{lan97,lan98}. The model we
shall use is defined in section 3 wherein various condensates
concerned are precisely defined. The effective gluon-exchange
interaction that sets the order of magnitude of the parameters 
of the model is also
given there. The phase structure that is expected to emerge as
density and/or temperature is increased is described in section 4.
Section 5 contains discussions. Notations and definitions -- some
of which are quite standard -- are given in the appendices to make
our discussions self-contained.

\section{ Phase transitions at finite chemical potential }
\label{sec:pt} 
\vskip 0.3cm

To render the paper self-contained and to clear up some of the
confusion on the key points of refs.\cite{lan97,lan98}, we briefly
contrast the ISB scenario of chiral restoration\footnote{There 
seems to be some confusion caused by the
term ``ISB" which to the best of our knowledge has not been used before
(apart from us)
in the literature. Since we do not know any other way of calling it, 
we shall continue using this term although it could very well be a
misnomer. The state of matter involved here was defined in ref.\cite{lan97}
and clearly illustrated in a toy condensed-matter model in \cite{lan98}.
One obvious question to ask is whether or not by ``ISB" we are talking about
a phase transition which is completely different from the chiral phase 
transition that is believed to be present in QCD. We should 
stress that the answer is no. That is, what we call ``ISB" would be 
implicitly present if QCD were solved reliably (see e.g., fig. 2
in \cite{berges}). What we are doing here -- which we believe is novel --
is exploiting a characteristic of chiral
phase transition that has not been identified before in the modeling of
the QCD phase structure. As defined below, the ``ISB" is characterized 
by the ``jump" in baryon density as one goes across the critical chemical
potential $\mu_c$ corresponding to chiral phase transition and our 
point is that this is driven by an attraction in the vector channel
caused by certain collective modes, in analogy to the potential introduced 
in the toy model of \cite{lan98} (see figures 1 and 2). We will argue
that this is related to the mechanism that may be
responsible for ``dropping" vector meson masses.} 
to the more standard mechanism as provided by a
constituent quark model.

\vskip 0.3cm

 In the last decade, numerous
investigations~\cite{klevanski} of constituent quark models have
shown that the constituent quark mass $M(\mu )$ rapidly decreases
if the chemical potential $\mu $ exceeds a certain value which we
denote by $\mu_c$. In the constituent particle picture the density
of states at a given momentum $p$, i.e.
\be
n(p) \; = \; \frac{1}{ \exp \left[ (E(p) - \mu )/T \right] +1 } \; ,
\hbo E(p) \; = \; \sqrt{ M(\mu )^2 \, + \, p^2 } \; ,
\label{eq:i1}
\en
is governed by a sharp Fermi surface at zero temperature. 
In this case, the density of constituent quarks remains zero for
$\mu \le \mu _{onset} = M(\mu )$. For $M(\mu )< \mu < \mu _c $, the density
is a monotonic increasing function of $\mu $. At the chiral phase transition
at $\mu = \mu _c$, the constituent mass drops. An inspection of (\ref{eq:i1})
shows that one expects a rapid increase of density.

\vskip 0.3cm
At finite density, our assertion is that collective modes due to
vibrations of the Fermi surface -- which must be present on a
general ground when the renormalization group flow to the Fermi surface
is properly studied in the effective Lagrangian or in hadronic models
of collective excitations\cite{KRBR} -- must be
taken into account for a correct description of finite density
hadron matter. In~\cite{lan98,lan97}, these modes were considered 
in terms of a local (effective) vector field $V_\mu $, the
vacuum expectation value (VEV) of the zeroth component of which
measures the departure from the constant chemical potential $\mu $
and the discontinuity in the Fermion number density. If the
chemical potential reaches a certain strength $\mu
_{ci}$, which is assumed to be smaller than $\mu _c$, an
expectation value $\langle V_0 \rangle $ forms and signals an
ISB (on top of the explicit
breaking via the chemical potential). In the constituent quark
model, the density of states is given in this case by
\be
n(p) \; = \; \frac{1}{ \exp \left[ (E(p) - \langle V_0 \rangle -\mu )/T
\right] +1 } \; .
\label{eq:i2}
\en
With the assumption $\langle V_0 \rangle + \mu_{ci} > \mu _c$ the
latter equation implies that the rise of the condensate $\langle
V_0 \rangle$ catalyzes the onset of the chiral phase transition.
The state of matter after chiral restoration is, in the present
case, considerably richer than what is given by the naive
(standard) scenario: as explained in~\cite{lan97}, due to the
dynamical origin of the condensate $\langle V_0 \rangle $, light
(non-relativistic) vector excitations are predicted to exist as
pseudo-Goldstone-type modes which dominantly decay into dilepton.
An intriguing possibility is that this light mode is just the lower
branch of the two-level model of \cite{KRBR}.

\vskip 0.3cm
Due to asymptotic freedom, there are two important deviations in
our effective quark theory from the naive constituent quark picture
which was used above. Note that the low energy effective quark model which we
shall employ below implements asymptotic freedom by a sharp
O(4)-invariant momentum cutoff. This analogy is of course only of
qualitative nature and may not be quite realistic on a quantitative
level.

\vskip 0.3cm
Firstly, the finite range of momentum integration implies a smeared
Fermi surface even at zero temperature. 
Including a self-consistent treatment of the field
$V_0$, small values of density might be observed also for $\mu <
\mu _{onset}$. Technically, this rise of density as function of
$\mu < \mu _{onset}$ is a consequence of the smeared Fermi
surface~\footnote{The behavior of density vs. chemical potential
for a low chemical potential $\mu << \mu_{ci}$ is irrelevant to the
issue in question. For a dilute system, the correct degrees of
freedom are nucleons or rather quasi-nucleons and {\it not} quarks
or quasi-quarks. There is no reason to believe that an effective
quark field theory without proper confinement mechanism can say
anything about that regime which is of no concern to this work.}.
{}From a physical point of view, this rise emerges from a
polarization of the Dirac sea. Similar effects were observed before
in NJL-type quark models when describing nucleons as solitonic
excitations~\cite{herbert}. 

\vskip 0.3cm
Secondly, the next question one might raise is
whether this quark system suffers from an instability signaled by a
steady increase of $\langle V_0
\rangle $. This instability is avoided by the shutdown of the quark
interaction at large momentum transfer as expected from asymptotic
freedom. This can be easily anticipated from the effective one-loop
potential for $V_0$ when one resorts to a quark model with sharp
momentum cutoff $\Lambda $. At least for values $V_0 > \Lambda $,
one observes a monotonic increase of the potential as function of
$V_0$ telling us that the formation of large values of $V_0$ is
unlikely to take place.

\vskip 0.3cm
Finally, the quark interaction mediated by the ``collective" field
$V_\mu $ is attractive (as suggested in \cite{KRBR})  
and thus favors the formation of the $V_0$
condensate at sufficiently strong values of the chemical potential.
One might question whether such an interaction is not an ad hoc
input foreign to the expected properties of low-energy QCD merely
to induce an ISB mechanism. The answer to this question is this: 
firstly, at least for $\vert V_0 \vert < \mu $, the collective field 
describes perturbations of the Fermi surface. These fluctuations 
account for new important physics which comes into play at finite densities, 
and which is present besides the microscopic interaction mediated 
by gluons. At the hadronic level, 
the schematic model of \cite{KRBR}
suggests that these collective modes could be
excitations of the $N^\star$-hole type and other varieties -- analogous
to phonons in solids and giant resonances in nuclei -- 
in the isoscalar vector channel. 
The exchange of such collective modes between quarks could contribute
an additional attraction.
It may be possible to determine the size of the contribution from the
collective modes using a model of the type studied in ref.\cite{KRBR}. 
But this program has not yet been fully worked out.
In what follows, since we do not really know how to pin down the
strength of the net attraction, we shall vary it within the range
indicated by the one-gluon exchange plus that of collective modes
and explore the qualitative structure of the possible phases.

\section{ From the model to the quark ground state }
\vskip 0.3cm

\subsection{The model}

We shall use a path-integral technique, so we define our model in
Euclidean space in order to ensure a proper convergence of the
functional integrals. In particular, the quark fields will be
Euclidean spinors. Our notation and conventions can be found in
appendix \ref{app:a}. The partition function which we will use
throughout this paper is given by
\bea
Z &=& \int {\cal D} q \;
{\cal D} \bar{q} \; {\cal D} q_c \; {\cal D} \bar{q}_c \; {\cal D}
\sigma \; {\cal D} \pi^a \; {\cal D} \Delta \; {\cal D} \Delta
^\dagger \; {\cal D} V_\mu \; \exp \left\{ \int d^4x \; L \right\} \; ,
\label{eq:2.1} \\
L &=& - \left( \matrix{ \bar{q} \cr \bar{q}_c }
\right) \, \left( \matrix{ S_0^{-1} & \kappa \Delta
^\dagger \cr \kappa \Delta & \widetilde{S}_0^{-1} } \right) \,
\left( \matrix{ q \cr q_c } \right) \; - \; g_\Delta \, \Delta
^\dagger \Delta
\label{eq:2.2} \\
&-&  V_\mu (x) \left( - \partial
^2 \delta _{\mu \nu } + \partial _\mu \partial _\nu \right) V_\nu
(x) \; - \; m^2_v \, V_\mu V_\mu \; - \; g_\chi \left[ (\sigma -m
)^2 \; + \; \pi ^2 \right] \; , \nonumber 
\ena 
where \bea S_0^{-1}
&=& \left[ i \dslash \, - \, \sigma (x) \, + \, i \gamma _5
\vec{\pi} (x) \vec{\tau} \, + \, i V_\mu (x) \gamma ^\mu \, + \, i
\, \mu \, \gamma ^0 \, \right] \; 1_{color} \; ,
\label{eq:2.3} \\
\widetilde{S}_0^{-1} &:=& \left[ i \dslash \, - \, \sigma (x) \, +
\, i \gamma _5 \vec{\pi} (x) \vec{\tau}^T \, - \, i V_\mu (x)
\gamma ^\mu \, - \, i \, \mu \, \gamma ^0 \, \right] \; 1_{color}
\; ,
\label{eq:2.4}
\ena
where $m$ is the current quark mass, $\mu
$ the chemical potential and $g_\Delta $, $g_\chi $ and $m_v$ are
(necessarily positive) coupling constants which control the
strength of the four-quark interactions that result when the bosonic
fields -- that we shall simply refer to as 
``mesonic" fields -- $\sigma $, $\vec{\pi}$, $V_\mu $ and $\Delta $,
$\Delta ^\dagger $ are integrated out. The role these couplings
play will be discussed below. The parameter $\kappa \in \{1,i\}$
will be chosen to reproduce the correct sign of the four-quark
current-current interaction which is well known for a correct
description of the low energy physics of the light mesons~\cite{ebe86}
(see subsection \ref{sec:2.3}).
The charge conjugated quark fields
possess the same kinetic energy as the quark fields and couple to
the flavor singlet vector fields with opposite sign~\cite{bai84}.
The quark fields $q$, $\bar{q} $ and $q_c$, $\bar{q}_c $ are
treated as independent fields. At the classical level, the
equations of motion yield the familiar relations
$\bar{q}=q^\dagger $ and $q_c = C (q^\dagger )^T$ for Euclidean
quark spinors (see appendix \ref{app:b}), where $C$ is the charge
conjugation matrix.

\vskip 0.3cm
Equation (\ref{eq:2.2}) is the simplest Lagrangian that contains all
relevant degrees of freedom for the physics we are interested in.
The quarks form flavor doublets and belong to the
fundamental representation of the SU(3) color group. The
``mesonic" fields $\sigma $, $\vec{\pi}$ and $V_\mu $ are color
singlets, $\sigma $ and $V_\mu $ are flavor singlets while the
pion fields form a flavor triplet. The quantum numbers of the
$\Delta $ field will be given below. The fields $\sigma $ and $\pi
^a$ are related to scalar and pionic degrees of freedom,
respectively. The (collective) 
vector field $V_\mu $ describes vibrations of
the Fermi surfaces~\footnote{As defined in (\ref{eq:2.3}) and (\ref{eq:2.4}),
the collective field $V_\mu\sim (V_0, \vec{0})$ carries constants that
reflect the (possibly density-dependent) strength of the coupling. They are
buried in the constant $m_v^2$ in (\ref{eq:2.2}). One could think of
the factor $m_v^2$
that we shall vary later reflecting, say, the $N^\star N^{-1}$-vector 
coupling strength of \cite{KRBR}.} 
and can be interpreted as the relativistic
analog of the zero sound or the giant 
dipole resonance~\cite{lan98}. Finally, $\Delta $ is the
conjugate variable coupled to the diquark composite field and can
therefore be viewed as the color superconductivity gap.

\vskip 0.3cm Let us study the symmetry of the model (\ref{eq:2.1})
in the case $\Delta , \, \Delta ^\dagger =0$. In the chiral limit
$m \rightarrow 0$, the Lagrangian $L$ for the two light-quark system,
(\ref{eq:2.2}), is invariant under an $SU(2)_A \times SU(2)_V$
flavor transformation of the quark fields $q$ and $q_c$,
respectively (see appendix \ref{app:c}). For a vanishing chemical
potential $\mu $, the model is O(4) invariant corresponding to the
Lorentz symmetry in Minkowski space. For $\mu \not=0$, the O(4)
symmetry is explicitly broken to a residual O(3) invariance.

\vskip 0.3cm
In order to compare the meson-induced interaction of the model
(\ref{eq:2.1}) with the
low energy effective quark interactions,
it is instructive to integrate out all meson fields. Defining
\be
\Delta =: \Gamma \delta
\label{eq:2.5}
\en
where $\delta $ is a c-number and $\Gamma $ specifies the Lorentz-,
flavor-  and color-structure of the superconducting gap, the following
quark interactions (in derivative expansion and
at low momentum transfer) are obtained
\bea
&+& m \, \bar{q}q \; + \;
\frac{1}{4 g _\chi } \left[
\bar{q}q \, \bar{q} q \; - \; \bar{q} \gamma _5 \tau ^a q \,
\bar{q} \gamma _5 \tau ^a q \, \right]
\; - \; \frac{1}{4 m_v^2} \, \bar{q} \gamma ^\mu q \, \bar{q}
\gamma ^\mu q
\label{eq:2.6} \\
&+& m \, \bar{q}_c q_c \; + \;
\frac{1}{4 g _\chi } \left[\bar{q}_c q_c \, \bar{q}_c  q_c \; - \;
\bar{q}_c \gamma _5 \tau ^a q_c \, \bar{q}_c \gamma _5 \tau ^a q_c \,
\right] \; - \; \frac{1}{4 m_v^2} \, \bar{q}_c \gamma ^\mu q_c \,
\bar{q} _c \gamma ^\mu q_c
\nonumber \\
&+& \kappa ^2 \, \frac{1}{g_\Delta } \, \bar{q}_c \Gamma q \, \bar{q}
\Gamma ^\dagger q_c \; .
\label{eq:2.7}
\ena

\subsection{ From the effective potential to the quark condensates }\label{2.2}

The effective potential is a convenient tool for studying the
ground state properties of an effective quark theory. Its global
minima determine the condensates or ground-state expectation values
of the bilinear quark fields that dictate the symmetries at low
energies. The procedure is quite general and, in the case we are
interested in, is equivalent to the BCS or Nambu-Gorkov formalism.

\vskip 0.3cm Let us be more specific. Firstly we derive a
non-linear meson theory by integrating out the quark fields. Note
that there is an intrinsic relation between quark condensates and
ground state expectation values of meson fields. Exploiting the
fact that the partition function (\ref{eq:2.1}) is invariant under
a shift of the mesonic integration variables yields the desired
relations
\bea
2 g_\chi \, \langle \sigma -m \rangle &=& \langle
\bar{q} q \rangle \; + \; \langle \bar{q}_c q_c \rangle
\nonumber \\
2 m_v^2 \, \langle V_0 \rangle &=& \langle i \bar{q} \gamma ^0
q \rangle \; - \; \langle i \bar{q}_c \gamma ^0 q_c \rangle
\nonumber \\
- \, 2 g_\Delta \, \langle \delta \rangle &=& \langle
\bar{q}_c \, \kappa \Gamma \, q \rangle \; + \; \langle \bar{q}
\, \kappa \Gamma ^\dagger \, q_c \rangle \; .
\nonumber
\ena
As expected, the quark and conjugate quark fields contribute to the
scalar condensate with equal signs while they come with opposite
signs in the
case of the density. $\langle \sigma \rangle $ serves as an order
parameter for the spontaneous breaking of chiral symmetry, while
$\langle \delta \rangle $ can be directly interpreted as the gap
of a color superconducting state. $\langle V_0 \rangle $ is
proportional to the quark density which is always non-zero for
non-vanishing chemical potential.

\vskip 0.3cm To define a convenient {indicator} -- say, a ``litmus paper"
-- $v_{ISB}$ which signals the onset of the ISB phase transition,
we first note that, in the limit $m_v^2 \rightarrow \infty$, the
four-quark interaction in the vector channel vanishes, and $V_\mu
$ is constrained to zero. The collective modes corresponding to the
fluctuations of the field $V_\mu $ decouple. In this case, the
density is entirely generated by the chemical potential. For finite
values of $m_v^2$, the interactions cause $V_0$ to depart from
zero. We define the {\it indicator} or signal for the ISB phase by
\be
\rho_{ISB} \; := \; \rho \bigg( \langle \sigma \rangle ,
\langle \delta \rangle , \langle V_0 \rangle \bigg) \; - \; \rho
\bigg( \langle \sigma \rangle , \langle \delta \rangle , 0 \bigg) \; ,
\label{eq:2.7b}
\en
i.e. we subtract from the full ground state density $\rho \left(
\langle \sigma \rangle , \langle \delta \rangle , \langle V_0
\rangle \right)$ the density \break $\rho \left( \langle \sigma
\rangle , \langle \delta \rangle , 0 \right)$ which is induced
solely by the chemical potential. $\rho _{ISB} $ therefore
directly measures the contribution of the collective dynamics
to the density and is equal to zero for $m_v^2
\rightarrow \infty $. While $\rho \left( \langle \sigma \rangle ,
\langle \delta \rangle , 0 \right)$ is expected to behave smoothly
under increase of the chemical potential, $\rho _{ISB}$ summarizes
the non-trivial behavior due to the ISB phase
transition.

\vskip 0.3cm To access the meson expectation values (in some
approximation), one first introduces external sources that
linearly couple to the meson fields in which we are interested.
The generating functional in the case of our model (\ref{eq:2.1})
is therefore 
\bea Z[j] &=& \int {\cal D} q \; {\cal D} \bar{q} \;
{\cal D} q_c \; {\cal D} \bar{q}_c \; {\cal D} \sigma \; {\cal D}
\pi^a \; {\cal D} \Delta \; {\cal D} \Delta ^\dagger \; 
{\cal D} V_\mu \exp
\left\{ \int d^4x \; [L + L _{source}] \right\} \; ,
\label{eq:2.10} \\ 
L_{source} &=& j_{\sigma } (x) \, \sigma (x) \;
+ \; j_0 (x) \, V_0 (x) \; + \;  j_{\Delta}(x) \, \delta (x) \; .
\label{eq:2.11} 
\ena 
The meson expectation value $\langle V_0
\rangle $, and therefore the condensate $\langle \bar{q} \gamma ^0
q \rangle $ will be non-trivial for non-vanishing values of the
chemical potential $\mu $. For convenience, we introduce the
compact notation $j=(j_{\sigma }, j_0, j_{\Delta} )$ and $\phi =
(\sigma , V_0, \delta )$. The effective action is defined by a
Legendre transformation with respect to the external fields
\be
{\cal A } [\phi _c] \; = \; - \, \ln Z[j] \; + \; \int d^4x \, j(x)
\phi _c(x) \; ,
\label{eq:2.12}
\en
where
\be
\phi _{cl} (x) \; := \; \frac{ \delta \, \ln Z[j] }{ \delta j(x) } \; .
\label{eq:2.13}
\en
The effective potential $U_{eff}(\phi_{cl})$ is obtained from
${\cal A}[\phi _{cl}]$
by resorting to space-time constant classical fields.

\vskip 0.3cm
In the case of many quark degrees of freedom (e.g. large number of colors),
the functional integral (\ref{eq:2.10}) can be approximately calculated
neglecting mesonic fluctuations of the mesons (including $\Delta $)
around their mean-field values. Within this approximation, we find
\bea
U_{eff}(\phi _{cl}) &=&
- \int \frac{d^4k }{(2\pi )^4 } \, \tr \, \ln
\left( \matrix{ \kslash - \sigma _{cl} + i \left(\mu + v_{cl}\right)
\gamma ^0 & \kappa \, \Gamma ^\dagger \delta _{cl} \cr \kappa \,
\Gamma \delta _{cl} &
\kslash - \sigma _{cl} - i \left( \mu + v_{cl}\right)
\gamma ^0 } \right)
\nonumber \\
&+& g_\chi \, (\sigma _{cl} -m )^2 \; + \; m_v^2 \, v_{cl}^2 \; + \;
g_\Delta \, \delta _{cl}^2 \; .
\label{eq:2.14}
\ena
It can be proven that the fermion determinant is real if the
diquark vertex $\Gamma $ is either hermitian or anti-hermitian.
This will be always the case for the examples studied below.
The classical fields at which the effective potential $U_{eff}$
takes its global minimum yield the desired condensates.
In the following, we will interpret $\sigma ^{min}_{cl} $ as quark condensate
and $v^{min}_{cl} $ as quark density, and
we will drop the index $cl$,
since  $U_{eff}$ consists only of classical
fields. We will study the phase structure of our model
(\ref{eq:2.1}) by referring to different types of Lorentz-
and color structures $\Gamma $ of the diquark condensate.

\subsection{ Gluon-induced quark interactions}
\label{sec:2.3}

In this subsection, we will estimate the size of the parameters
that figure in our model by
comparing (\ref{eq:2.6},\ref{eq:2.7}) with the low-energy quark
theory which may be obtained from effective one-gluon
exchange. Note however that the resulting four-Fermi interaction
represents a lot more physics than the exchange of one bare or dressed
gluon and the effective size of the parameters will
be fixed only when the additional corrections that arise in the 
decimation toward the Fermi surface (coming from,
e.g., significant collective
modes including the $N^\star N^{-1}$ excitation of \cite{KRBR}) 
are suitably accounted for.

\vskip 0.3cm Quark interactions which are mediated by the exchange
of a dressed gluon can be described in terms of the action
\bea
S_{1gluon} &=& - \int d^4x \; \bar{q} \left[ i \dslash \, - \,
A^a_\mu \, T^a \, \gamma ^\mu \right] q \, - \, \int d^4x \;
\bar{q}_c \left[ i \dslash \, + \, A^a_\mu \, \left(T^a \right)^T
\, \gamma ^\mu \right] q _c
\nonumber \\
&-& \int d^4x \;
d^4y \; A^a_\mu (x) \, D^{ab}_{\mu \nu }(x-y) \, A^b_\nu (y) \; ,
\label{eq:2.20}
\ena
where $T^a$ are the eight generators of the SU(3)
color group. If $\Gamma $ denotes the quark gluon vertex, the
charge conjugate quark fields couple to the gluons via $C \Gamma
^T C^{-1}$~\cite{bai84}. Much work has been done on investigating
the properties of light hadrons resorting to a wide variety
 of
ans\"atze for the dressed effective gluon propagator $D^{ab}_{\mu
\nu }(x-y)$~\cite{fi82,alk88,pa77}. Note that $D^{ab}_{\mu \nu }$
is assumed to be a positive definite operator
to ensure convergence of the gluonic functional integral. When
employed in the Dyson-Schwinger equation for ground state properties
and in the Bethe-Salpeter equation for description of light
mesons, these effective interactions are known to provide a
successful description of the spontaneous breakdown of chiral
symmetry and the corresponding low-energy pion physics. Over the
last decade, research has extended these effective
1-gluon-exchange models by using non-trivial quark-gluon vertices
that follow from Taylor-Slavnov identities. These investigations have
met with a remarkable success in low-energy hadron
phenomenology~\cite{sme91,rob94}.

\vskip 0.3cm Here, we shall not resort to such
Dyson-Schwinger-type approaches to color superconductivity, but
will employ the corresponding low energy effective quark theory
which emerges from an elimination of the high energy modes by means
of a renormalization group approach. In such an approach, the
effective four-quark contact term provided by the current-current
quark interaction naturally emerges, since this interaction respects
all symmetries of low energy QCD. Moreover, it was observed in
strongly coupled QED that such an interaction plays an important
role in low energy physics due to a large anomalous dimension~\cite{bar89}
of the corresponding operator.
In the context of low energy QCD, the current-current interaction
dictates the Lorentz, color and flavor structure while its strength
must be chosen to correctly describe the physics of the light
mesons~\cite{ebe86}.
Such an effective low-energy quark theory
coming from a renormalizable QCD-inspired quark model generically
yields a correct phase structure although it may fail in the
description of certain dynamical processes (for details
see~\cite{lanjl}).

\vskip 0.3cm
When applying the renormalization group decimation of energy scales
to the model (\ref{eq:2.20}), the current-current interaction which
is compatible with the low energy QCD symmetries is given by
\bea
L &=& \ldots + \, G \, \bigl[ \; \bar{q} \, 1_F T^A \gamma
^\mu \, q \; \bar{q} \, 1_F T^A \gamma ^\mu \, q \; + \; \bar{q}_c
\, 1_F \left(T^A \right)^T\gamma ^\mu \, q_c \; \bar{q}_c \, 1_F
\left(T^A\right)^T \gamma ^\mu \, q_c
\nonumber \\
&-& 2 \,\bar{q}_c \, 1_F \left(T^A\right)^T \gamma ^\mu \, q_c \;
\bar{q} \, 1_F T^A \gamma ^\mu \, q \; \bigr] \; ,
\label{eq:2.21}
\ena
where its absolute strength $G$ must be chosen for e.g. reproducing
the correct pion physics in free space.

\vskip 0.3cm
We follow the ideas first presented in~\cite{al97} for the diquark
phase, and assume that a diquark condensate respects chiral
symmetry while it breaks the residual global color symmetry from
SU(3) down to SU(2). The color, flavor and Dirac structure
of the diquark vertices which we will investigate below
are (see appendix \ref{app:b})

\vskip 0.3cm
\begin{tabular}{ll}
\hspace{3cm} $ \Gamma \; = \; \epsilon _{ik} \,
\epsilon ^{\alpha \beta 3 } \, \gamma _5 $ \hspace{3cm} & (scalar) \cr
\hspace{3cm} $ \Gamma \; = \; \epsilon _{ik} \, \epsilon
^{\alpha \beta 3 } \, $ \hspace{3cm} & (pseudo-scalar) \cr
\hspace{3cm} $ \Gamma \; = \; \epsilon _{ik} \, \epsilon ^{\alpha
\beta 3 } \, \gamma ^\mu \gamma _5 $ \hspace{3cm} & (vector) \cr
\hspace{3cm} $ \Gamma \; = \; \epsilon _{ik} \, \epsilon ^{\alpha
\beta 3 } \, \gamma ^\mu $ \hspace{3cm} & (axial-vector),
\end{tabular}

\vskip 0.3cm
where $i,k = 1,2 $ act in flavor space, and $\alpha , \beta
= 1 \ldots 3$ are color indices.
Equation (\ref{eq:2.21}) will be our starting point
for estimating the coupling strengths of our model (\ref{eq:2.1}).
For this purpose, we rearrange the four-quark contact interaction
(\ref{eq:2.21}) by means of a Fierz transformation so as  to
make sure that the classical equations of motion, i.e. the
mean-field level, of our model equal the Dyson-Schwinger equations
with the interaction (\ref{eq:2.21}). We relegate the details of
this calculation to appendix \ref{app:d}. The result is 
\bea L &=&
\ldots + \; \frac{G}{4} \left( \bar{q} \, 1_F 1_c \, q \; \bar{q}
\, 1_F 1_c \, q \; - \; \bar{q} \, 1_c\gamma _5 \tau ^a \, q \;
\bar{q} \, 1_c \gamma _5 \tau ^a \, q \, \right) 
\label{eq:2.22} \\ 
&-& G \left[ \frac{1}{2N}- \frac{1}{8} \right] \, 
\bar{q} \, 1_F 1_c \gamma ^\mu \, q \; \bar{q}
\, 1_F 1_c \, \gamma ^\mu q 
\nonumber \\ 
&+& \frac{G}{4N}
\left(\bar{q}_c\right)^\alpha _i \, \epsilon _{ik} \epsilon
^{\gamma \alpha \beta } \gamma ^\mu \, q^\beta _k \; \bar{q}
^\kappa _l \, \epsilon _{lm} \epsilon ^{\gamma \kappa \omega }
\gamma ^\mu \, \left(q_c\right)^\omega _m 
\nonumber \\ 
&+& \; \frac{G}{4N} \left(\bar{q}_c\right)^\alpha _i \, \epsilon _{ik}
\epsilon ^{\gamma \alpha \beta } \gamma ^\mu \gamma _5\, q^\beta
_k \; \bar{q} ^\kappa _l \, \epsilon _{lm} \epsilon ^{\gamma
\kappa \omega } \gamma ^\mu \gamma _5\, \left(q_c\right)^\omega _m
\nonumber \\ 
&+& \frac{G}{2N} \left(\bar{q}_c\right)^\alpha _i \,
\epsilon _{ik} \epsilon ^{\gamma \alpha \beta } \; q^\beta _k \;
\bar{q} ^\kappa _l \, \epsilon _{lm} \epsilon ^{\gamma \kappa
\omega } \; \left(q_c\right)^\omega _m 
\nonumber \\ 
&-& \frac{G}{2N} \left(\bar{q}_c\right)^\alpha _i \, \epsilon _{ik}
\epsilon ^{\gamma \alpha \beta } \, \gamma _5 \; q^\beta _k \;
\bar{q} ^\kappa _l \, \epsilon _{lm} \epsilon ^{\gamma \kappa
\omega } \, \gamma _5 \; \left(q_c\right)^\omega _m \; , 
\nonumber
\ena 
where $1_F, 1_c$ are unit matrices in flavor and color spaces,
respectively, and $G$ sets the scale of the interaction strength.
$N=3$ is the number of colors. Comparing this result with
(\ref{eq:2.6},\ref{eq:2.7}), we first deduce the ``sign'' of the
meson quark couplings, i.e.
\be
\kappa _{scalar} = i \; , \hbo
\kappa _{ps} = 1 \; , \hbo
\kappa _{vector} = i \; , \hbo
\kappa _{axial} = 1 \; ,
\label{eq:2.23}
\en
and then read off the corresponding strengths, i.e. \bea g_\chi
&=& \frac{1}{2G} \; , \hbo m_v^2 \; = \; \frac{N}{G} 
\frac{1}{1- N/4} \; ,
\label{eq:2.24} \\ 
g^{scalar}_\Delta &=& g^{ps}_\Delta \; = \;
\frac{2N}{G} \; , \hbo g^{vector}_\Delta \; = \; g^{axial}_\Delta
\; = \;  \frac{4N}{G} \; . 
\ena 
In particular, we find the following ratios for $N=3$
\be
g_\chi \; : \; m_v^2 \; : \; g^{scalar}_\Delta
\; : \; g^{vector}_\Delta \; = \; 1 \; : \; 6 \; : \; 6
\; : \; 12 \; .
\label{eq:2.24b}
\en
In the next section, we will study either case of diquark condensation
and will tacitly assume that a mixed phase, e.g. with the
scalar and vector diquark condensates simultaneously present,
will not occur.

\vskip 0.3cm
The  parameter ratios (\ref{eq:2.24b}) represent the quark 
interactions carrying the quantum numbers of one gluon exchange.
As explained in section 
\ref{sec:pt}, the attraction already present in the current-current
interaction for the $V_0$ channel with
three colors and two flavors (see Appendix
D for a caveat that may follow from a different Fierz rearrangement) 
is likely to be further enhanced by collective modes. In our
calculation, this effect will be duly accounted for 
by lowering the mass $m_v^2$. Since little is known at present
about the size of this additional attraction, we will 
explore in the next section the phase structure for a wide span of 
parameters while the order of magnitude of the parameters will be
kept at the values given by the effective one-gluon exchange.

\section{ Phase structure }\label{sec:3}
\vskip 0.3cm

At finite baryon density, a rich phase structure is expected to
appear in QCD. Whereas at zero density a quark condensate is
formed in the QCD quark sector, two possible phases might exist at
finite values of the chemical potential: the ISB
phase in which Lorentz symmetry is ``spontaneously" broken in the sense 
defined in \cite{lan97,lan98}
and the phase in which a diquark condensate is formed
which breaks the residual global color symmetry of
QCD~\cite{al97}. In the following subsections, we will explore the
phase structure which arises from our low-energy effective quark
theory when the chemical potential is increased.

\vskip 0.3cm
\subsection{ Scalar and pseudo-scalar diquark condensations }

We follow the ideas first presented in~\cite{al97} for the diquark
phase. There, it was conjectured that the color
superconductor instability occurs in the scalar diquark.
In this section, we will extend the analysis of~\cite{al97} to
the study of possible scalar, pseudo-scalar, vector and axial-vector diquark
condensations as well.

\vskip 0.3cm
In this subsection, we will study the possible
emergence of scalar as well as pseudo-scalar diquark condensates,
i.e.
\bea
&& \left\langle \bar{q}_{c \, i }^\alpha \; \gamma _5 \, \epsilon _{ik}
\, \epsilon ^{\alpha \beta 3} \; q_k^\beta \right\rangle
\hbox to 4cm{ \hfill (scalar) \hfill }
\label{eq:3.1} \\
&& \left\langle \bar{q}_{c \, i }^\alpha \; \epsilon _{ik} \,
\epsilon ^{\alpha \beta 3} \; q_k^\beta \right\rangle
\hbox to 4cm{ \hfill (pseudo-scalar) \hfill } \; .
\label{eq:3.2}
\ena
Here $i,k =1 \ldots 2$ are flavor indices, while $\alpha, \beta = 1
\ldots 3$ are color indices. Dirac indices are not explicitly
shown. The behavior of the Euclidean quark spinors under Lorentz
transformations can be found in appendix \ref{app:b}. If one of
these condensates is realized in a finite-density quark matter
while the quark condensate is zero,
chiral symmetry is restored (see appendix \ref{app:c2}).

\vskip 0.3cm We first consider the case of the pseudo-scalar
superconducting gap, $\Delta = \epsilon _{ik} \epsilon ^{\alpha
\beta 3} \, 1 \, \delta $ in conjunction with the ISB phase
transition extensively discussed in~\cite{lan97,lan98}. The
fermion determinant in (\ref{eq:2.14}) can be calculated in a
closed form (see appendix \ref{app:e}), i.e.
\bea \Det_{ps} &=& \prod _k \; \left( k_+^2 +
\sigma ^2 \right)^{12} \; \left( k_-^2 + \sigma ^2 \right)^4 \;
\nonumber \\ && \left[ \left( w_+ k_0 - i(\mu +v) \, w_-
\right)^2 \, + \, w_+^2 \vec{k}^2 \, + \, w_-^2 \sigma ^2
\right]^8\label{eq:3.3}
\ena
where $$ k_\pm  \; := \; (k_0 \pm i (\mu+v),
\vec{k} )^T, \hbo w_\pm  \; := \; 1 \, \pm \, \frac{ \delta ^2 }{
k_+^2 + \sigma ^2 } \; . $$ The effective potential is \bea
U_{eff}^{ps} (\sigma  , \delta  , v ) &=& - \, \ln \; \Det _{ps}
\nonumber \\ &+& g_\chi \, (\sigma   -m )^2 \; + \; m_v^2 \,
v^2  \; + \; g_\Delta \, \delta  ^2 \; . \label{eq:3.4} \ena The model
(\ref{eq:2.1}) is considered to be an effective low-energy quark
theory in that the momentum integration implicit in (\ref{eq:3.4})
is cut off at the scale $\Lambda $. Note that for a reliable
investigation of the model at finite values of the chemical
potential it is crucial that the momentum cutoff leave the
Lorentz, i.e. O(4), symmetry intact. To assure this, we used a
sharp O(4) invariant momentum cutoff.

\vskip 0.3cm In the case of the scalar diquark composite, we
choose $\Delta = \gamma _5 \epsilon _{ik} \epsilon ^{\alpha \beta
3 } \, \delta $. The functional determinant in (\ref{eq:2.14}) is
given by (see appendix \ref{app:e})
\bea \Det_{scalar} &=& \prod _k \; \left( k_+^2 + \sigma
^2 \right)^{12} \; \label{eq:3.10} \\ && \left( k_-^2 + \sigma ^2
\right)^4 \; \left[ \left( w_+ k_0 - i(\mu +v) \, w_- \right)^2 \,
+ \, w_+^2 \vec{k}^2 \, + \, w_+^2 \sigma ^2 \right]^8. \nonumber
\ena Comparing (\ref{eq:3.10}) with (\ref{eq:3.3}), one
immediately sees that the determinants with pseudo-scalar and
scalar diquark entries, respectively, coincide if chiral symmetry
is respected, i.e. $\sigma =0$. In the following, we will assume
that we will not encounter a strong CP-problem, but that the
scalar diquark condensation takes place.

\vskip 0.3cm We explored the parameter space, looking for the
global minimum of the effective potential $U_{eff}$, by resorting
to Monte Carlo techniques. We find generically three different
minima of the effective potential. One minimum shows a strong
chiral condensate while the superconducting gap and $v$ are close
to zero. We call this minimum $min -\sigma $. The second minimum,
so-called $min-v$, is located at strong values of $v$ with the
order parameters $\sigma $, $\delta $ being small. Finally, the
third minimum is dominated by a large value of the superconducting
gap $\delta $, and we shall call this $min-\Delta $. All these
minima are possible candidates for the true ground state which is
characterized by the {\it global } minimum of $U_{eff}$.

\begin{figure}[t]
\centerline{
\epsfxsize=7cm
\epsffile{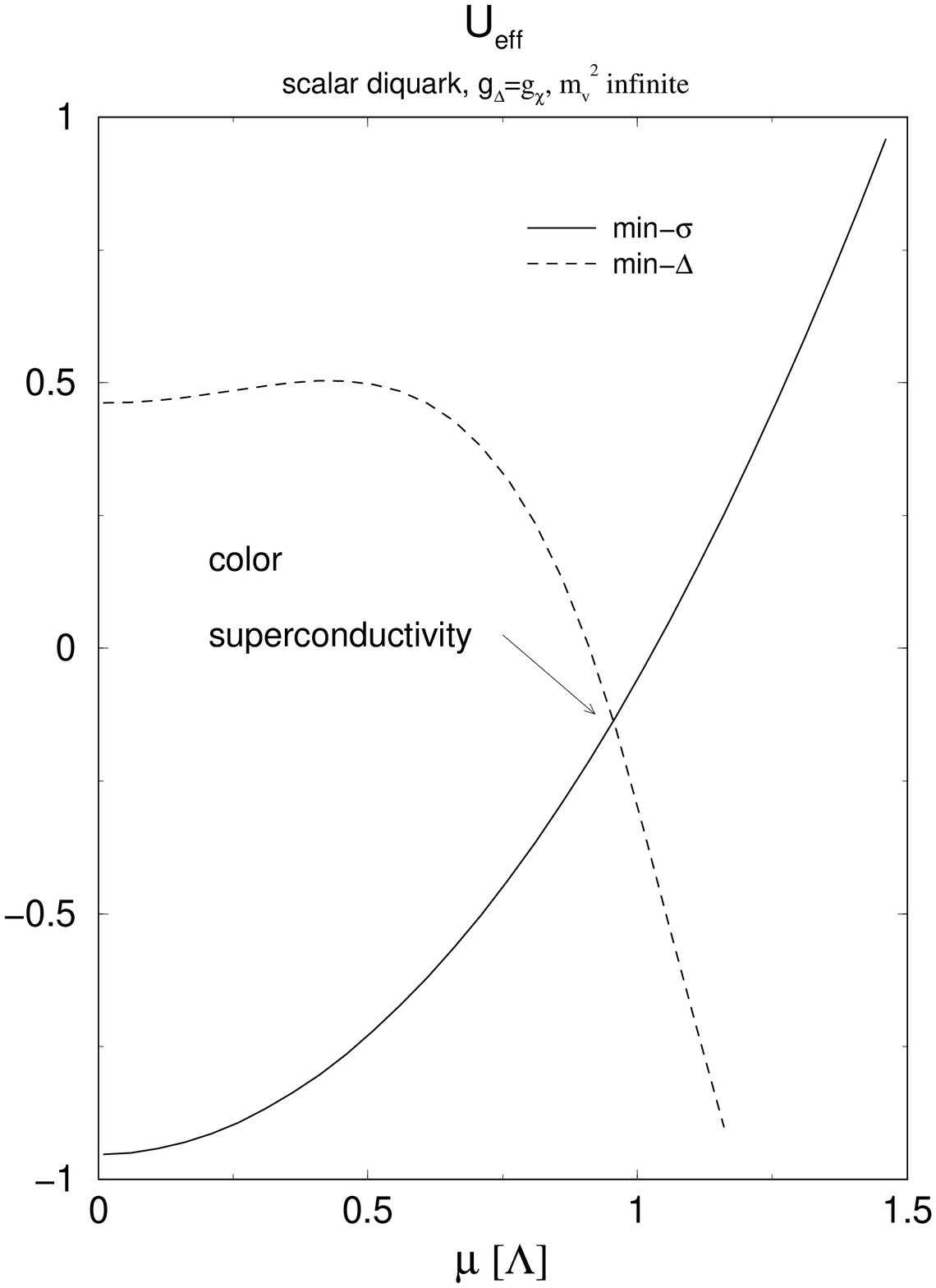}
\hspace{1cm}
\epsfxsize=7cm
\epsffile{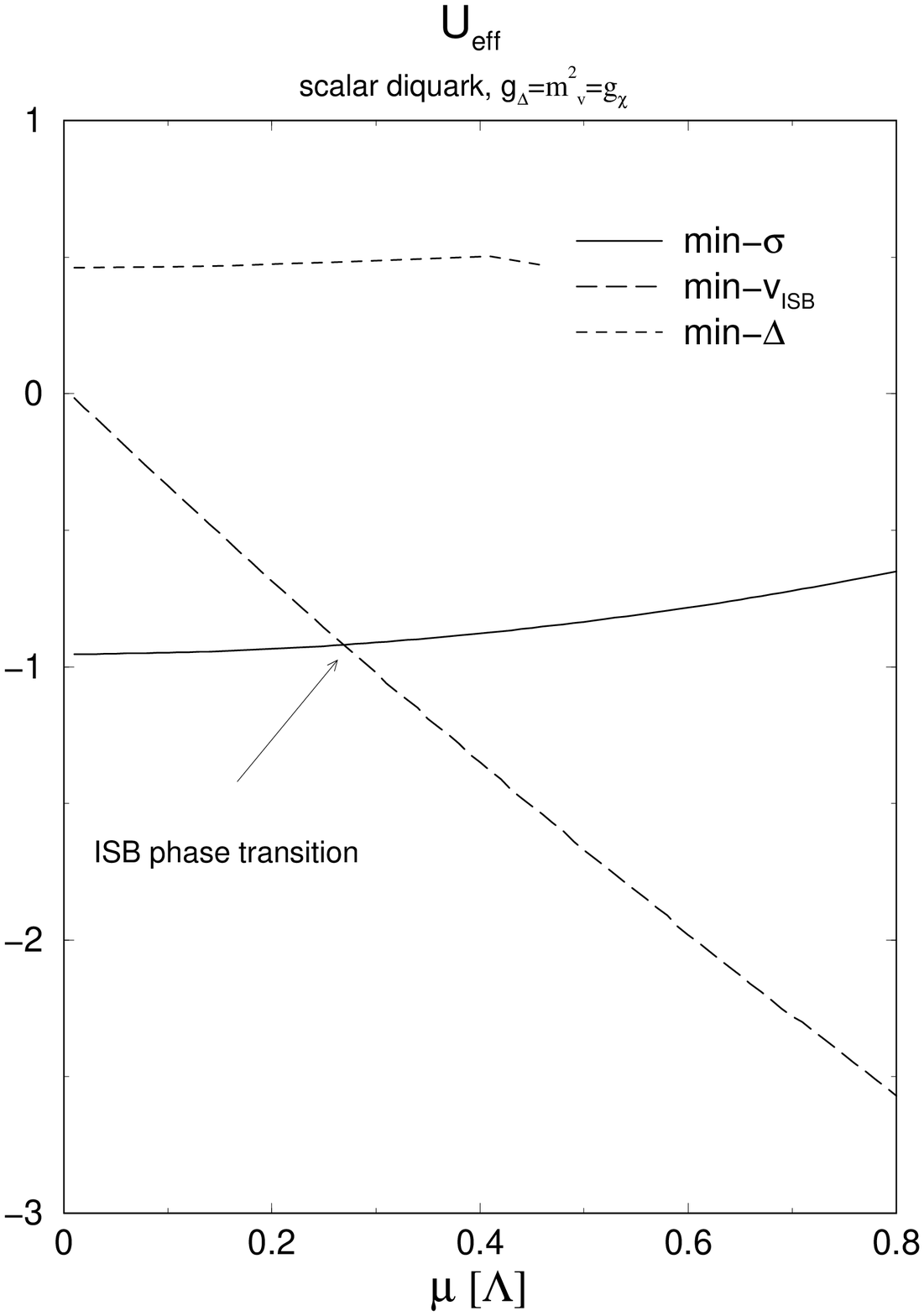}
}
\caption[]{ Scalar diquark case: the effective potential $U_{eff}$ in units
   of $\Lambda ^4 /4 \pi^2$ of the local minima as function
   of the chemical potential $\mu $ for $m_v^2 \rightarrow \infty $
   (left) and $m_v^2=g_\chi=g_\Delta $ (right). }\label{fig:2}
\end{figure}
\vskip 0.3cm To establish the instability due to color
superconductivity at high density as proposed in~\cite{al97}, we
turn off the interaction strength in the vector channel by tuning
$m_v^2 \rightarrow \infty $ and study the competition between the
chiral phase and the superconductor phase as a function of the
chemical potential. The numerical result of our model is shown in
figure \ref{fig:2} (left panel). It indicates that the scalar diquark
condensation sets in at some large values of the chemical
potential $\mu = {\cal O}(\Lambda )$, where the use of a
four-Fermi contact interaction is most likely unjustified. Since
we are here only interested in the qualitative phase structure, we
shall not consider the non-local interaction due to the full
effective one-gluon exchange.

\vskip 0.3cm The situation becomes more dramatic if we use a
moderate value for $m_v^2$. 
For instance for $g_\Delta = g_\chi = m_v^2 = \Lambda ^2 / 4\pi^2
$, the effective potential at the minima comes out to be as shown in
figure \ref{fig:2} (right panel).
It turns out that ISB phase transition takes
place at a small value ($\mu \approx 0.4 \Lambda $) of the
chemical potential, whereas the minimum which is characterized by
the scalar diquark condensate, i.e. $min-\Delta$, turns into a
saddle point.

\begin{figure}[t]
\centerline{
\epsfxsize=8cm
\epsffile{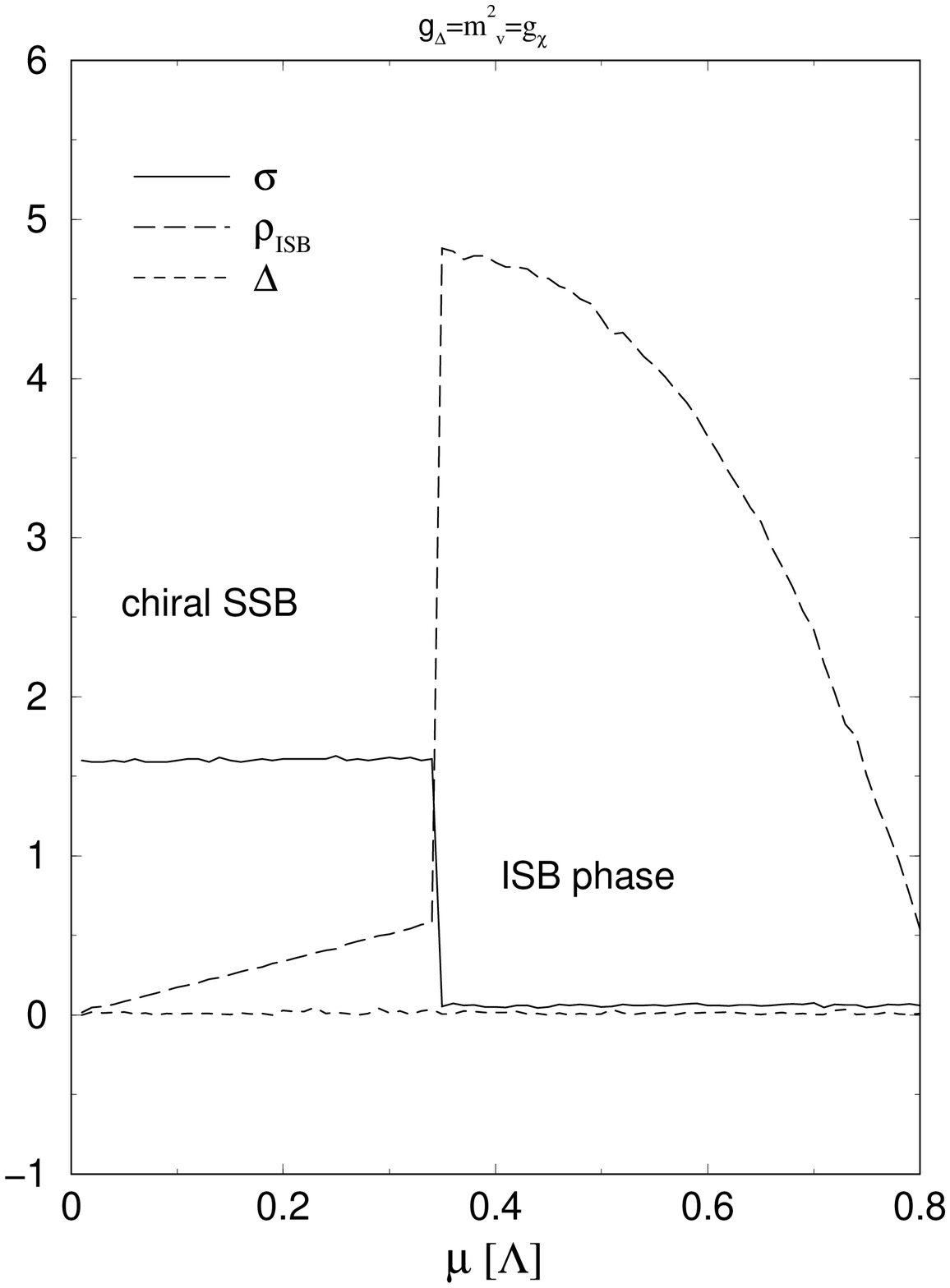}
\epsfxsize=8cm
\epsffile{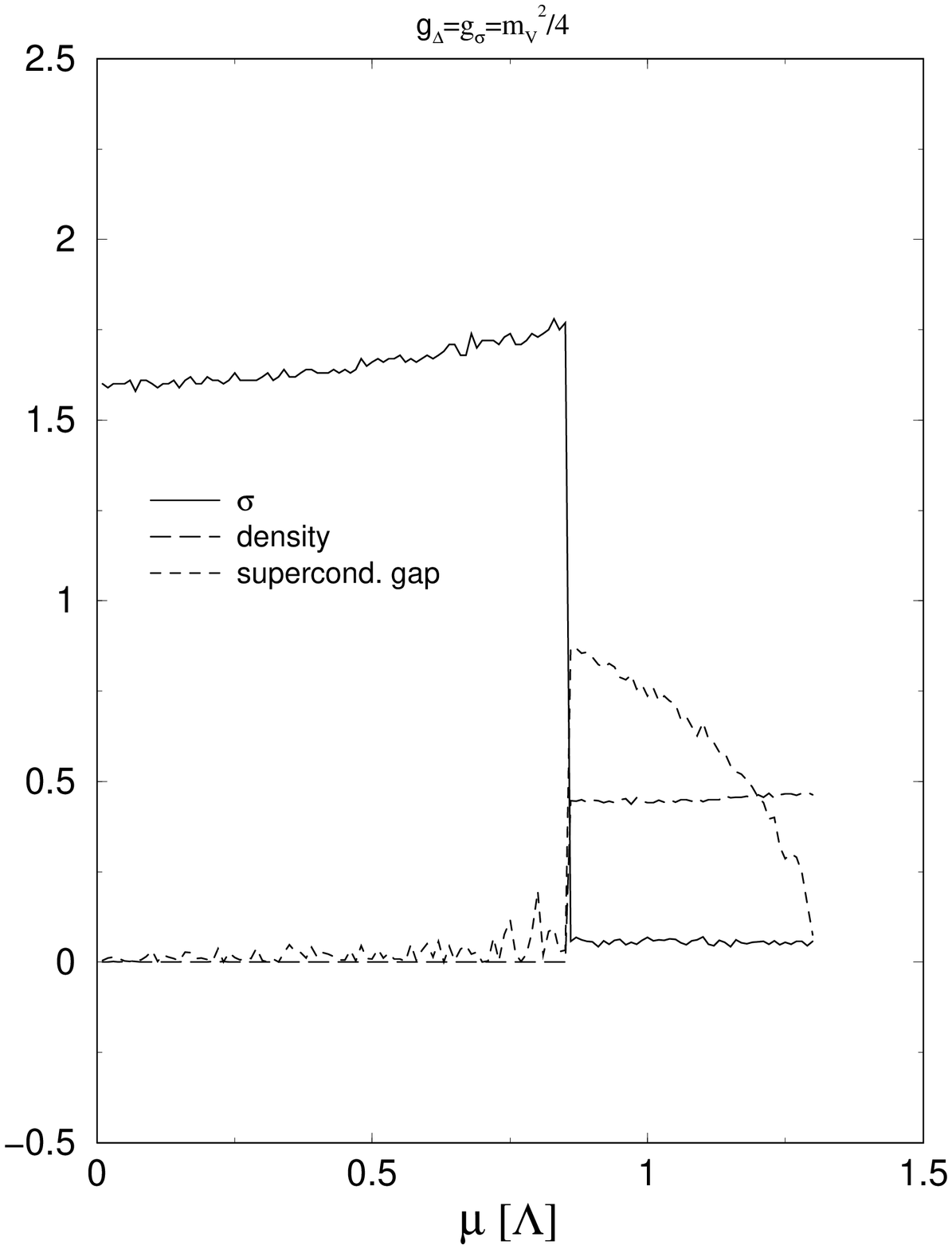}
}
\caption{ The chiral condensate $\sigma $, ISB indicator $\rho _{ISB}$ and
   superconducting gap $\delta $ as function of the chemical potential
   for the 1-gluon exchange (left) and an unnatural weak
   four-quark interaction
   in the vector channel (right). }\label{fig:4}
\end{figure}
\vskip 0.3cm In the case of finite $m_v^2$, we resorted to a full
Monte-Carlo treatment to calculate the strength of the order
parameters $\sigma $, $v$ and $\delta _{vec}$. It turns out that
the scalar diquark condensate is not important for ground state
properties if we choose the typical interaction strength of order unity,
i.e. $g_\chi = g_\Delta = m_v^2 = \Lambda ^2 /4\pi^2 $ 
(see
figure \ref{fig:4}). At zero chemical potential, the quark
condensate $(\sigma   \not=0 )$ signals the spontaneous breakdown
of chiral symmetry. The ISB phase ($min-v$) and the
superconducting phase ($min-\Delta$) appear as meta-stable states.
When the chemical potential is increased above a certain critical
value $\mu $, a first-order phase transition takes place from the
chiral broken phase to the ISB phase. An ``induced" breaking of
Lorentz symmetry takes place while chiral symmetry gets restored.
This mechanism as well as its consequences for the light particle
spectrum were extensively discussed in~\cite{lan97,lan98}.

\vskip 0.3cm We then reduced the strength of the quark
interactions in the vector channel by a factor of $4$, i.e.
$g_\chi = g_\Delta = \Lambda ^2 /4\pi^2 $, $m_v^2 = 4 \, \Lambda
^2 /4\pi^2 $, as this is the channel that gives rise to the ISB
phase. In fact, we searched for a region of parameter space where
the superconducting state constitutes the state with the lowest
energy density. In this case, one indeed observes a
superconducting state at large values of the chemical potential
(see figure \ref{fig:4}). It appears, however, that if one takes
the interaction strength to be of order unity -- which is 
a typical strength one expects as before,
the color superconductivity is unlikely to play a role in the
density regime relevant for the physics we are interested in.
This is because the ISB phase has a lower vacuum energy
density.

\vskip 0.3cm
\subsection{ Vector diquark condensation }

\begin{figure}[t]
\centerline{
\epsfxsize=8cm
\epsffile{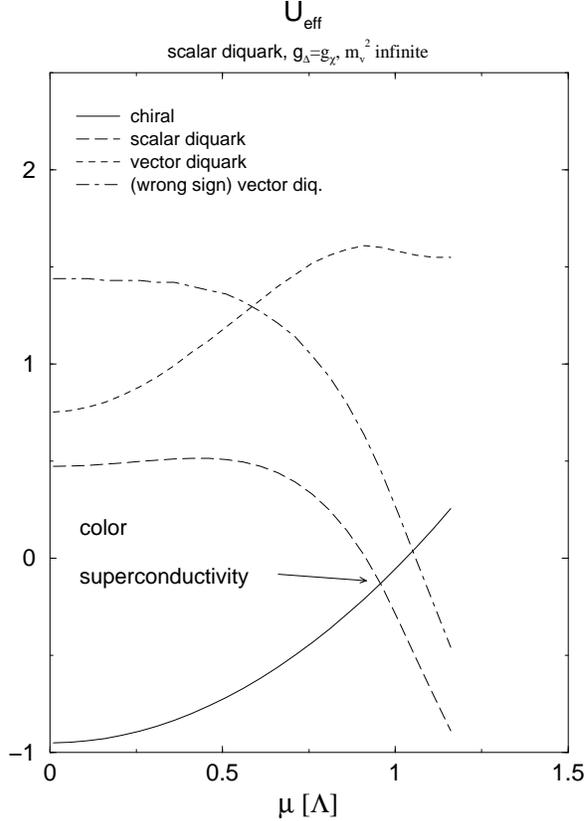}
}
\caption{ The effective potential $U_{eff}$ in units of
   $\Lambda ^4 /4 \pi^2$ of the local minima as function
   of the chemical potential $\mu $ for $m_v^2 \rightarrow \infty $.
   }\label{fig:3}
\end{figure}
One might wonder whether a vector diquark condensate and/or an
axial-vector diquark condensate, i.e. \bea && \left\langle
\bar{q}_{C \, i }^\alpha \; \epsilon _{ik} \, \epsilon ^{\alpha
\beta 3 } \, \gamma ^\mu \, \gamma _5 \; q ^\beta _k \,
\right\rangle \; , \hbox to 4cm{ \hfill (vector) \hfill }
\label{eq:3.20} \\ && \left\langle \bar{q}_{C \, i }^\alpha \;
\epsilon _{ik} \, \epsilon ^{\alpha \beta 3 } \, \gamma ^\mu \; q
^\beta _k \, \right\rangle \; , \hbox to 4cm{ \hfill
(axial-vector) \hfill } \; , \label{eq:3.20b} \ena might not
compete with the ISB phase transition if the interaction is given
by the effective one-gluon exchange. Since both diquark
condensates transform as vectors, one might suspect that
they could be more sensitive to the chemical potential which
transforms as the fourth component of the vector.

\vskip 0.3cm The fermion determinant in (\ref{eq:2.14}) can be
readily calculated for the vector diquark ($\Delta = \epsilon
_{ik} \epsilon ^{\alpha \beta 3} \, \gamma ^0 \, \gamma _5 \,
\delta $) and the axial-vector ($\Delta = \epsilon _{ik} \epsilon
^{\alpha \beta 3} \, \gamma ^0 \, \gamma _5 \, \delta $) diquark
(see appendix \ref{app:e}),
\bea \Det_{vec} &=& \prod _k \; \left( k_+^2 + \sigma ^2
\right)^{12} \; \left( k_-^2 + \sigma ^2 \right)^4 \;
\label{eq:3.22} \\ && \left[ \left( w_- k_0 - i(\mu +v) \, w_+
\right)^2 \, + \, w_+^2 \vec{k}^2 \, + \, w_-^2 \sigma ^2
\right]^8 \nonumber \\ \Det_{axial} &=& \prod _k \; \left( k_+^2 +
\sigma ^2 \right)^{12} \; \left( k_-^2 + \sigma ^2 \right)^4 \;
\label{eq:3.22b} \\ && \left[ \left( w_- k_0 - i(\mu +v) \, w_+
\right)^2 \, + \, w_+^2 \vec{k}^2 \, + \, w_+^2 \sigma ^2
\right]^8 \nonumber \ena In order to gain some insight into the
competition between the chiral phase and the phases exhibiting the
various diquark condensates, we studied the limit $m_v^2
\rightarrow \infty$ which amounts to the suppression of the ISB
mode. The effective potential of the various local minima -- which are
the possible candidates for the true ground state -- is shown in
figure \ref{fig:3} as function of the chemical potential. It turns
out that the scalar diquark condensation sets in at $\mu = {\cal
O}(\Lambda )$ whereas the other diquark condensates merely
correspond to meta-stable states.

\vskip 0.3cm The reason why the vector diquark condensates have no
significant influence on the ground state properties is that the four-quark
interaction in the vector diquark channel has the ``wrong'' sign.
If for the purpose of illustration one changes this sign by
setting $\kappa^2 _{vector }=1$ (compare with (\ref{eq:2.23})), the
vector diquark order parameter has a much stronger effect (see
figure \ref{fig:3}), but still cannot prevent diquarks from
condensing in the scalar channel.

\subsection{ Finite temperatures }

Finally, we study the impact of temperature on the magnitude of
the order parameters $\sigma $, $\Delta $ and the ISB indicator
$\rho _{ISB}$. In view of the results of the previous subsections,
we will only focus on the emergence of the scalar diquark
condensate.

\vskip 0.3cm In taking into account temperature effects, we use
the imaginary time formalism by introducing Matsubara frequencies
and by replacing the $k_0$ momentum integration in (\ref{eq:3.22})
by a discrete sum, i.e.
\be
k_0 \; = \; (2n+1) \, \pi \, T \; , \hbo
\int dk_0 \rightarrow 2\pi T \, \sum _n \; ,
\label{eq:3.50}
\en
where $T$ is the temperature. Note that the momentum integration
is truncated by an O(4) symmetric momentum cutoff $\Lambda $, i.e.
$k_0^2 + \vec{k}^2 \leq \Lambda ^2 $. This implies that the
momentum integral vanishes for $\pi T > \Lambda $. From a physical
point of view, this amounts to assuming that the non-local (e.g.
gluon-induced ) interactions at large momentum transfer ($k^2 >
\Lambda ^2$) that match the four-Fermi interaction of our
effective quark theory at the scale $\Lambda $ contributes only
perturbatively  to the order parameters $\sigma $, $v$ and $\Delta
$ (see~\cite{lan96}). We therefore expect the melting of all
condensates at $T_c \approx \Lambda / \pi $. Using a generic
energy scale of $\Lambda \approx 600 \, $MeV, one finds roughly
$T_c \approx 200 \, $MeV (at zero density), which is the right
order of magnitude as revealed by lattice QCD
studies~\cite{kan97}.

\begin{figure}[t]
\centerline{
\epsfxsize=8cm
\epsffile{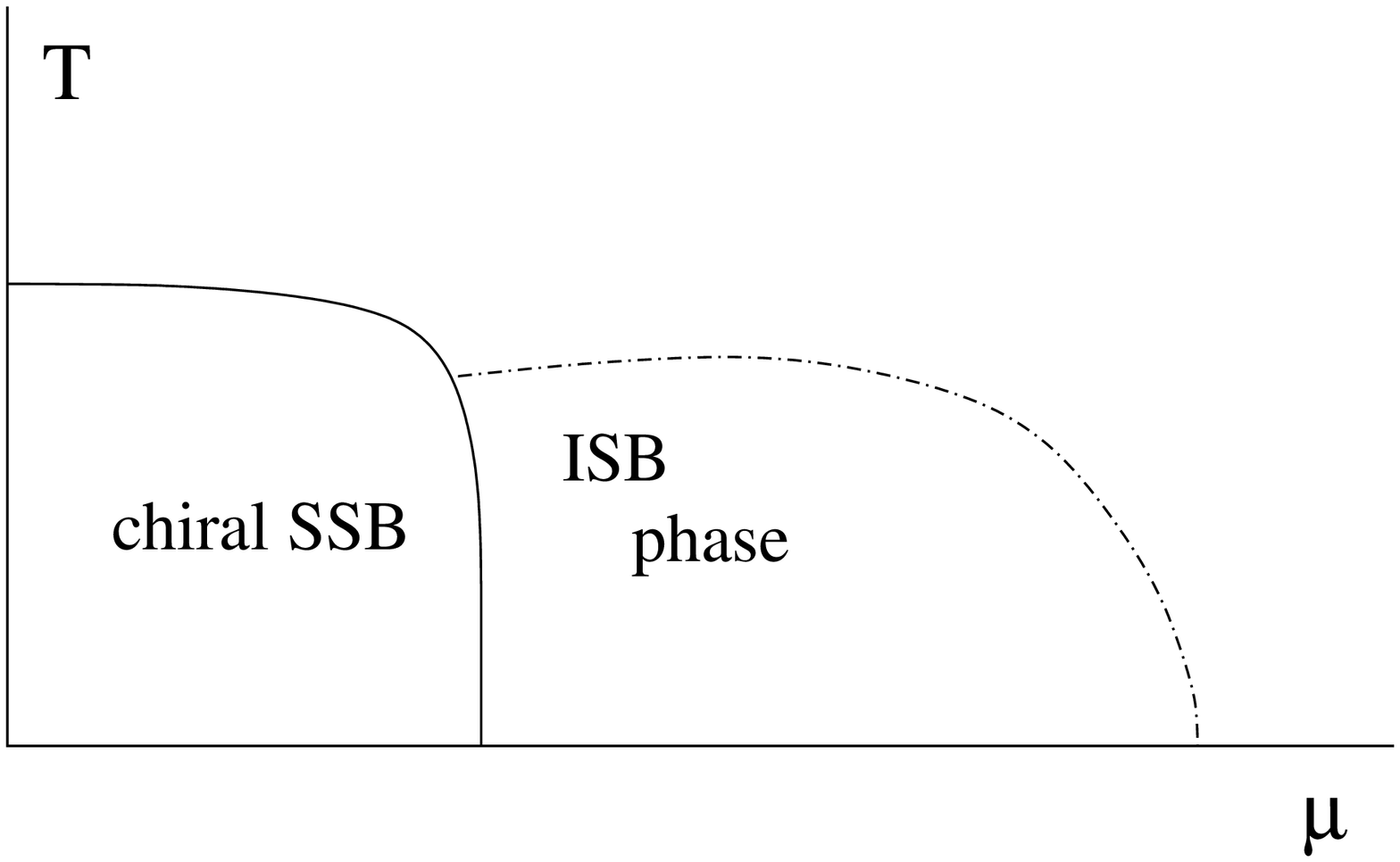}
\epsfxsize=8cm
\epsffile{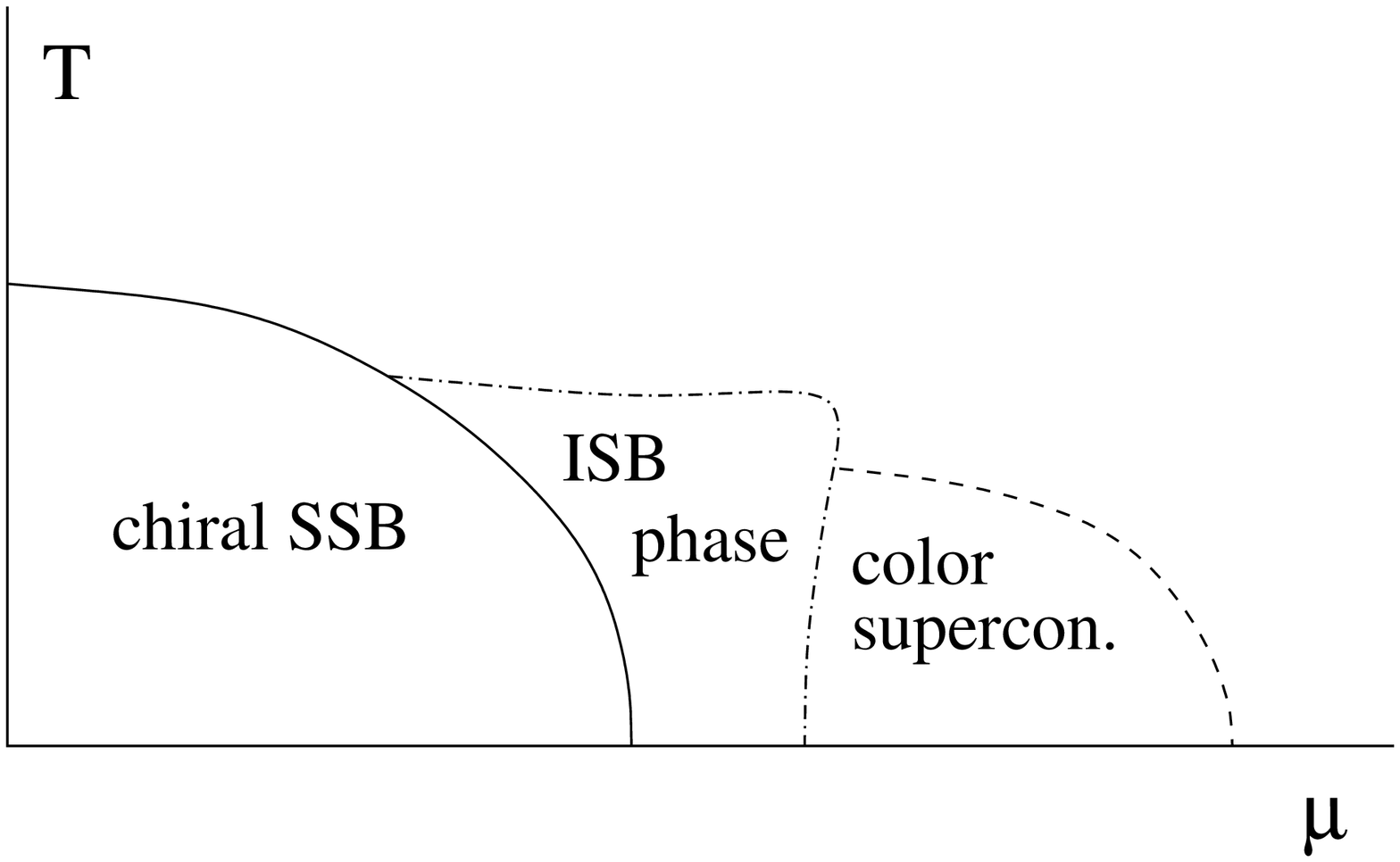}
}
\caption{ Phase diagram for a medium size $m_v^2=\Lambda ^2/4\pi^2$
   interaction strength in the vector channel (left) and for a
   small size interaction $m_v^2=8 \Lambda ^2/4\pi^2$ (right).
   }\label{fig:6}
\end{figure}
\vskip 0.3cm Using Monte Carlo techniques, we searched for
a given chemical potential and a given temperature for the global
minimum of the effective potential $U_{eff}$ which occurs for the
parameters $\sigma $, $v$ and $\delta $. We then calculated
$\rho_{ISB}$ (\ref{eq:2.7b}) for the signal of the ISB phase
transition. Even at a medium scale of the chemical potential ($\mu
< \Lambda $) one observes a significant jump of $\rho _{ISB}$.
Although, strictly speaking, there is no phase transition but a
crossover, one can define rather sharp phase boundaries in
parameter space by \vskip 0.3cm
\begin{tabular}{lll}
\hbox to 3cm {\hfill } &
$\sigma > v_{ISB}$, $\sigma > \delta $ & chiral phase \cr
& $v_{ISB} > \sigma $, $v_{ISB} > \delta $ & ISB phase \cr
& $\delta > \sigma$, $\delta > v_{ISB} $ & color superconductor phase, \cr
\end{tabular}

\vskip 0.3cm
where $v_{ISB} := \rho _{ISB}/2m_v^2$.
The numerical result of our model is sketched in figure
\ref{fig:6}.

\vskip 0.3cm One finds that for the interaction strengths $g_\chi
$, $g_\Delta ^{scalar}$, $m_v^2$ all of order unity, diquark
condensation plays no role. It is the ISB phase $\rho _{ISB}
\not=0 $ that characterizes the quark state. At large values of
the chemical potential, we expect the contribution from the quark
interactions to the density to become small, and that there is a
smooth transition from the ISB phase at some medium values of the
chemical potential to the phase of perturbative QCD where the
density is largely induced by the chemical potential.

\vskip 0.3cm If one were to choose an unnaturally weak vector
interaction strength \hfill \break
$m_v^2 \gg g_\chi , g_\Delta ^{scalar}$, one
might observe ``shells'' built up from the ISB phase and the color
superconducting phase, respectively. We note, however, that the
parameter range $\mu \approx \Lambda $ is beyond the scope of the
present model. In order to access these ranges of the chemical
potential, a non-local and asymptotically free quark interaction
(as proposed e.g. in~\cite{rob94}) might be required.

\section{ Discussions }
\label{sec:dis}

A detailed analysis of the possible phase structure of finite
density quark matter based on an effective low-energy quark theory
is presented. The four-quark contact interaction used in the study
is provided by the effective
quark current-current interaction which naturally
emerges from a renormalization group decimation of high energy
scales in a many-body system with a Fermi sea. 
From a phenomenological point of view,
the effective current-current interaction is well known for a correct
description of the physics of the light mesons at least at low densities.
The simplest way to understand what we have done in this paper is that our
strategy corresponds to interpolating
the {\it effective}
bottom-up partition function (\ref{eq:2.1}) with the explicit
{\it collective fields} $V_\mu$ and $\Delta$ 
(as well as the pseudoscalars $\pi$
and the scalar $\sigma$) to that with the four-Fermi interactions 
(\ref{eq:2.21}) that must
arise from a top-down approach starting from the QCD partition function.
Although the effective theory might be incomplete in some aspects of
nonperturbative structure of QCD, such as lack of quark
confinement and additional many-body effects, 
we believe that it qualitatively describes the phases
of quark matter when the chemical potential is varied. Hopefully
such an approximate description of quark matter phases at high
density will serve as an important input for a thorough
renormalization group study (based on a decimation of energy
scales relative to the Fermi momentum~\cite{pol92}) which will
then yield more quantitative results.

\vskip 0.3cm Our finding is that the generic interaction strength
provided by the current-current interaction in the diquark channel
is much too weak to trigger a scalar diquark
condensate. Pseudo-scalar, vector and axial-vector diquark
condensates are even less favored. Instead of the color
superconductivity, we find an induced Lorentz symmetry breaking to
occur at medium values of the chemical potential leading to
chiral symmetry restoration with color symmetry remaining
unbroken.

\vskip 0.3cm In sum, we are led to the following conjecture for
the quark phase structure: at zero density, strong QCD
interactions break chiral symmetry spontaneously leading to quark
condensation. At small finite values of the chemical potential,
the quark condensation weakens in the presence of
Lorentz-symmetry-breaking terms, and a ``roll-over'' from $\langle
\bar{q} q \rangle $ to $\langle \bar{q} \gamma _0 q \rangle $ is
observed (called ``ISB phase transition"). {One of our main
points  
is that this phenomenon in quark language is related to the hadronic
description of the chiral phase transition and the ``dropping mass"
of the $\omega$ meson in \cite{KRBR}.}
As the chemical
potential is further increased, the quark interaction is weakened
due to asymptotic freedom. At a certain (second) critical value of
the chemical potential, the interaction is no longer strong enough
to sustain quark condensation in one of the Lorentz channels. In
this case, it is most likely that color superconductivity occurs
since an arbitrarily weak (but attractive) interaction at the
Fermi surface is sufficient for (color) Cooper pair formation.

\vskip 0.3cm 
We should mention that our description may be much too simplisitic
to be realistic. It has been shown by Ilieva and Thirring \cite{thirring}
that even in a simple quantum mechanical system  
involving both the ISB-like (called ``mean field" by the authors of
\cite{thirring}) phase and superconducting phase, the phase structure can be
quite intricate and complicated. It is perhaps significant to note in
particular that
as in the Ilieva-Thirring model, an attractive ``mean field" potential
could even trigger (instead of prevent)
superconductivity in contrast to what we have found in this paper. It is 
fair to say that the situation in QCD could be immensely more intricate
and complex.

\vskip 0.3cm
Finally let us comment on the apparent breaking of
the local gauge invariance by the formation of the diquark
condensate. In a recent important work, Schaden et
al.~\cite{sch98} exploited a topological field theory to obtain a
complete gauge fixing while avoiding the Gribov problem. In the
so-called equivariant gauge, gauge invariance was fixed up to a
{\it global } symmetry in color space. The authors then showed
that quark confinement (in these gauges only) could be induced by
a spontaneous breakdown of this residual global color symmetry. As
a consequence, owing to Goldstone's theorem, light colored fields
(which arise in these gauges) decouple from local gauge invariant
operators but contribute to the Wilson loop (due to its non-local
nature)~\cite{sch98}. These Goldstone bosons then could produce an
area law (and therefore confinement) due to inherent infra-red
singularities~\cite{sch98}.

\vskip 0.3cm Except on a lattice, any approach to color
superconductivity, be that via an effective quark theory or via an
renormalization group analysis, will have to be based on a
completely gauge fixed QCD. In this respect, diquark condensation
{\it does not} break local gauge invariance, but breaks instead
the residual global color symmetry. In view of the results
presented in~\cite{sch98}, one encounters the interesting
possibility that a hypothetical phase transition to a color
superconducting state at high density restores chiral symmetry,
but does not imply quark deconfinement. However, we should recall
that, at least in certain supersymmetric Yang-Mills theories,
chiral restoration and deconfinement occur
simultaneously~\cite{sei94}.

\vspace{1cm}
\centerline{\bf Acknowledgments: \hfill }

\vskip 0.3cm We have benefited from valuable discussions with
Gerry Brown. Both of us are grateful to Korea Institute for
Advanced Study where this work was done
for hospitality and to C.W. Kim, Director of the
Institute, for supporting and encouraging an exploratory
collaboration in astro-hadron physics at KIAS. KL would like to
thank H.~Reinhardt for encouragement and support.

\vspace{1cm}

\appendix
\setcounter{equation}{0}\renewcommand{\theequation}{\mbox{A.\arabic{equation}}}
\section{ Notation and conventions }
\label{app:a}

The metric tensor in Minkowski space is
\be
g_{\mu \nu } \; = \; \hbox{diag} ( 1 , -1 , -1 , -1 ) \; .
\label{eq:a1}
\en
We define Euclidean tensors $T_{(E)}$ from the tensors in Minkowski
space $T_{(M)}$ by
\be
T^{\mu _1 \ldots \mu _N }_{ (E) \phantom{\ldots \mu _N } \nu _1
\ldots \nu _n }
\; = \; (i)^r \, (-i)^s \;
T^{\mu _1 \ldots \mu _N }_{ (M) \phantom{\ldots \mu _N } \nu _1
\ldots \nu _n } \; ,
\label{eq:a2}
\en
where $r$ and $s$ are the numbers of zeros within $\{ \mu _1 \ldots
\mu _N \}$ and $\{ \nu _1 \ldots \nu _n \} $, respectively.
In particular, we have for the Euclidean time and the Euclidean
metric
\be
x_{(E)} ^ 0 \; = \; i \, x_{(M)}^0 \; , \hbo
g^{\mu \nu }_{(E)} \; = \; \hbox{diag} ( -1 ,-1 ,-1 ,-1 ) \; .
\label{eq:a3}
\en
Covariant and contra-variant vectors in Euclidean space differ by an
overall sign. For a consistent treatment of the symmetries, one is forced
to consider the $\gamma ^\mu $ matrices as vectors. Therefore, one is
naturally led to anti-hermitian Euclidean matrices via (\ref{eq:a2}),
\be
\gamma ^0 _{(E)} = i \gamma ^0 _{(M)} \; , \hbo
\gamma ^k _{(E)} = \gamma ^k _{(M)} \; .
\label{eq:a4}
\en
In particular, one finds
\be
\left( \gamma ^\mu _{(E)} \right) ^{\dagger } \; = \;
- \; \gamma ^\mu _{(E)} \; , \hbo
\{ \gamma ^\mu _{(E)} , \gamma ^\nu _{(E)} \} \; = \;
2 g^{\mu \nu }_{(E)} \; = \; -2 \, \delta _{\mu \nu } \; .
\label{eq:a5}
\en
The so-called Wick rotation is performed by considering the
Euclidean tensors (\ref{eq:a2}) as real fields.

\vskip 0.3cm
In addition, we define the square of an Euclidean
vector field, e.g. $V_\mu $, by
\be
V^2 \; := \; V_\mu V_\mu \; = \; - V_\mu V^\mu \; .
\label{eq:a51}
\en
This implies that $V^2$ is always a positive quantity (after the wick
rotation to Euclidean space).

\vskip 0.3cm
The Euclidean action $S_E$ is defined from the action $S_M$ in Minkowski
space by
\be
\exp \{ i S_M \} \; = \; \exp \{ S_E \} \; .
\label{eq:a52}
\en
Using (\ref{eq:a2}), it is obvious that the Euclidean Lagrangian $L_E$
is obtained from the Lagrangian $L_M$ in Minkowski space by replacing
the fields in Minkowski space by Euclidean fields, i.e.
$L_E = L_M$.

\vskip 0.3cm
Let the tensor $\Lambda ^{\mu }_{\phantom{\mu } \nu } $ denote a
Lorentz transformation in Minkowski space, i.e.
\be
\Lambda ^{\mu }_{\phantom{\mu } \alpha } \Lambda ^{\nu }_{\phantom{\nu }
\beta } \, g^{\alpha \beta } \; = \; g^{\mu \nu } \; .
\label{eq:a6}
\en
Using the definition (\ref{eq:a2}), one easily verifies that
$\Lambda ^{\mu }_{(E) \, \nu } $ are elements of an O(4) group, i.e.
$ \Lambda ^{T}_{(E)} \Lambda _{(E)} = 1$, which is the counterpart
of the Lorentz group in Euclidean space.
\vskip 0.3cm

In order to define the Euclidean quark fields, we exploit the
spinor transformation of the Euclidean quark field
\bea
q_{(E)}(x_E) \rightarrow q^\prime _{(E)}(x^\prime _E) &=&
S(\Lambda _{(E)}) q_{(E)}(x_E) \; , \hbo
x_E \rightarrow x^\prime _E \; = \; \Lambda x_E \; ,
\label{eq:a7} \\
S(\Lambda _{(E)}) \gamma ^\mu _{(E)} S^\dagger (\Lambda _{(E)})
&=& \left( \Lambda ^{-1} _{(E) } \right)^{\mu }_
{\phantom{\mu }
\nu } \, \gamma ^\nu _{(E)} \; ,
\label{eq:a8}
\ena
where the matrices are given by
\be
S(\Lambda _{(E)}) \; = \; \exp \{ - \frac{i}{4} \omega _{\mu \nu }
\sigma ^{\mu \nu }_{(E)} \}
\label{eq:a81}
\en
for $\dete(\Lambda)=1$ only. It is obvious that one must interpret
\be
\bar{q}_{(E)} \; = \; q^\dagger _{(E)}
\label{eq:a9}
\en
in order to ensure that e.g.~the quantity $\bar{q}_{(E)} \gamma ^\mu _{(E)}
q_{(E)} $ transform as an Euclidean vector.
We suppress the index $E$ throughout the paper and mark tensors
with an index $M$, if they are Minkowskian.

\setcounter{equation}{0}\renewcommand{\theequation}{\mbox{B.\arabic{equation}}}
\section{ Charge conjugation in Euclidean space }
\label{app:b}

The equation of motion of a Dirac spinor which couples to a U(1)
gauge field is given by
\be
-i \partial _\mu \, q^\dagger (x) \, \gamma ^\mu \; - \; q^\dagger (x) \,
e \Aslash \; = \; 0 \; .
\label{eq:b1}
\en
Defining the charge conjugated spinor by
\be
q_c (x) \; := \; C \, \left( q^\dagger (x) \right) ^T
\label{eq:b2}
\en
with some Dirac matrix $C$ which will be specified below, equation
(\ref{eq:b1}) can be cast into
\be
C \, i \dslash ^T \, C^{-1} \; q_c(x) \; + \; e \, C \, \Aslash ^T \,
C^{-1} \; q_c(x) \; = \;
0 \; .
\label{eq:b3}
\en
One finds that $q_c(x)$ satisfies a Dirac equation with reversed charge
$e \rightarrow -e $ if $C$ satisfies $ C \gamma ^T _\mu C^{-1}
= - \gamma _\mu $. Let us choose a standard representation of the Dirac
matrices, i.e.
\be
\gamma ^0 = \left( \matrix{ i & 0 \cr 0 & -i } \right) \; ,
\hbo
\gamma ^k = \left( \matrix{ 0 & - \tau _k  \cr \tau_k & 0 }
\right) \; ,
\hbo
\gamma _5 = \left( \matrix{ 0 & 1  \cr 1 & 0 } \right)
\label{eq:b4}
\en
with $\tau _k$ the SU(2) Pauli matrices. Note that
\be
\{ \gamma ^\mu, \gamma _5 \} \; = \; 0 \; , \hbo \gamma _5^\dagger
\; = \; \gamma _5^T \; = \; \gamma _5 \; .
\label{eq:b5}
\en
Using this representation,
one finds
\be
C \, = \, \gamma ^0 \gamma ^2 \; , \hbo C^\dagger \, = \, C^{-1}
\, = \, C^T \, = \, -C \; .
\label{eq:b6}
\en
Let us investigate the behavior of the charge conjugated spinor $q_c$
under Lorentz, i.e. O(4), transformations (\ref{eq:a81}).
Replacing $S(\Lambda ) q$ for $q$ in the Dirac equation (\ref{eq:b1})
and re-deriving $q_c$ from this equation, one finds that the
charge conjugated spinors transform as
\be
q_c^\prime \; = \; C \, \left( S^{-1}(\Lambda ) \right)^T \,
C^{-1} \; q_c \; .
\label{eq:b7}
\en
We now confine ourselves to Lorentz transformations (\ref{eq:a81})
with $\dete(\Lambda)=1$ and will check the behavior of $q_c$ under
parity transformations later.
Firstly note that
$$
\left( \gamma _\mu \gamma _\nu - \gamma _\nu \gamma _\mu \right)^T
\; = \; \gamma _\nu^T \gamma _\mu^T - \gamma _\mu^T \gamma _\nu^T
\; = \; - C \; \left( \gamma _\mu \gamma _\nu - \gamma _\nu \gamma _\mu
\right) \, C^{-1} \; ,
$$
and therefore
\be
S^T(\Lambda ) \; = \; C \, S^\dagger (\Lambda ) \, C^{-1} \; ,
\hbo (\dete \Lambda =1) \; .
\label{eq:b8}
\en
Using the last equation, one readily verifies for a solution $q_c$
of the equation of motion that
$$
q_c^\prime \; = \; C \left( q^\dagger S^\dagger (\Lambda ) \right)^T
\; = \; C \left( S^\dagger \right)^T \, \left(q^\dagger \right)^T
\; = \; S(\Lambda ) \, q_c \; .
$$
An inspection of (\ref{eq:a8}) shows that $S(\Lambda ) = \gamma _0$
generates a reflection of spatial coordinates and therefore
serves as an example for a Lorentz transformation with
$\dete \Lambda = -1$. We learn from
(\ref{eq:b7}) that the conjugated spinor behaves as $q_c
\rightarrow - q_c$ under this reflection transformation. We
therefore conclude that the quark spinor $q_c$ transforms as
\be
q_c^\prime \; = \; \dete(\Lambda ) \, S(\Lambda ) \; q_c \; .
\label{eq:b9}
\en
In particular, one finds that the diquark condensates $\langle
\bar{q}_c \gamma _5 q \rangle $ and $\langle \bar{q}_c \gamma _\mu
\gamma _5 q \rangle $ transform as a scalar and a vector,
respectively, as in Minkowski-space.

\vskip 0.3cm
\setcounter{equation}{0}\renewcommand{\theequation}{\mbox{C.\arabic{equation}}}
\section{ Chiral symmetry in Euclidean space }
\label{app:c}

\vskip 0.3cm
\subsection{Generalities}

For simplicity, we will illustrate the emergence of chiral symmetry
in effective meson theories for the case of two quark flavors.
Generalization to $N_f$ quark flavors is obvious.
One observes that the quark kinetic energy, i.e.
\be
\bar{q} \left( i \dslash  \, - \, \sigma \, + \, i
\gamma _5 \, \vec{\pi } \vec{\tau } \right) q \;
\label{eq:c1}
\en
is invariant under a simultaneous transformation of meson and
quark fields. Let $V$ be an SU(2) matrix which acts only on flavor
indices, and ${\cal V}$ a matrix which exists in flavor and Dirac
space, i.e.
\be
V \; = \; \exp \left\{ i \vec{\alpha } \vec{\tau } \right\} \; , \hbo
{\cal V} \; = \; \exp \left\{ i \vec{\alpha } \vec{\tau } \, 1_D
\right\} \; ,
\label{eq:c2a}
\en
For convenience, we introduce the flavor matrix
\be
M \; = \; \sigma \, - \, i \vec{\pi } \vec{\tau } \; .
\label{eq:m}
\en
Then the invariance under ${\cal V} \in SU_V(2)$ transformation is
observed for
\be
q^\prime \; = \; {\cal V} \, q \; , \hbo \bar{q}^\prime \; = \; \bar{q}
\, {\cal V} ^\dagger \; , \hbo M^\prime \; = \; V \, M \, V^\dagger \; .
\label{eq:c2}
\en
To display the axial-vector symmetry, we choose $U$ to be another
SU(2) matrix acting in flavor space and define
\be
U \; = \; \exp \left\{ i \vec{\beta } \vec{\tau } \right\} \; ,
\hbo {\cal U} \; = \; \exp \left\{ i \vec{\beta } \vec{\tau } \,
\gamma _5 \right\} \; .\label{eq:c3a}
\en
Invariance under axial-vector chiral rotations, ${\cal U} \in
SU_A(2) $, is generated by
\be
q^\prime \; = \; {\cal U} \, q \; , \hbo \bar{q}^\prime \; = \; \bar{q}
\, {\cal U} \; , \hbo M^\prime \; = \; U \, M \, U \; .
\label{eq:c3}
\en
The so-called chiral circle is invariant separately under both
transformations, i.e.
\be
\sigma ^2 \, + \, \vec{\pi}^2 \; = \; \frac{1}{2} \tr \left(M M^\dagger
\right) \; .
\label{eq:c4b}
\en
Note, that $q$ and $\bar{q}$ are independent fields. This makes it possible
to assign different flavor matrices to either field.
Equations (\ref{eq:c2}) and (\ref{eq:c3}) establish the well known
$SU_A(2) \times SU_V(2)$ chiral symmetry.

\vskip 0.3cm
There is a subtle difference between the Euclidean formulation and the
formulation in Minkowski space. Let $\psi $ and $\bar{\psi}$  denote the
spinor solutions of the equations of motion. In Minkowski space, one
finds
\be
\bar{\psi}_{(M)} = \psi^\dagger _{(M)} \gamma ^0 _{(M)} \; , \hbo
\bar{\psi}^\prime _{(M)} = \left( {\cal U} \psi _{(M)} \right)^\dagger
\gamma ^0_{(M)} = \bar{\psi} _{(M)} \; {\cal U}
\label{eq:c5}
\en
Equation (\ref{eq:c5}) implies that the saddle points are degenerate
by axial chiral rotations. This situation is different in Euclidean
space. There one finds
\be
\bar{\psi}_{(E)} = \psi^\dagger _{(E)} \; , \hbo
\bar{\psi}^\prime _{(E)} = \left( {\cal U} \psi _{(E)} \right)^\dagger
\; \not= \; \bar{\psi} _{(E)} \; {\cal U}
\label{eq:c6}
\en
implying that the degeneracy of the saddle points is lifted. Since
we are interested in the non-linear meson theory that results from
integrating out the quark fields, the resulting fermion
determinant depends only on the chiral circle (\ref{eq:c4b}) in
either formulation.

\vskip 0.3cm To illustrate this point for the Euclidean case, we
calculate the fermionic determinant for the space-time constant
fields $\sigma $, $\vec{\pi}$. The eigenvalues of the Dirac
operator in momentum space, i.e. $\kslash - \sigma + i \vec{\pi}
\vec{\tau } \, \gamma _5 $, are given by $\lambda _{\pm } =
-\sigma \pm i \sqrt{ k^2 + \vec{\pi}^2 } $. One therefore obtains
the desired result
\be
Det \; = \; \prod _k (\lambda _+ \lambda _-)^2 \; = \;
\prod _k \left( k^2 + \sigma ^2 + \vec{\pi }^2 \right)^2 \; .
\label{eq:c7}
\en

\vskip 0.3cm
Finally, we must specify the behavior of the other independent quark
fields $q_c$, $\bar{q}_c$ under chiral rotations. We define
\bea
q^\prime _c &=& \left({\cal V}^\dagger \right)^T \, q_c \; , \hbo
\bar{q}_c^\prime \; = \; \bar{q}_c \, {\cal V}^T \;
\hbo {\cal V} \in SU_V(2) \; ,
\label{eq:c8} \\
q^\prime _c &=& {\cal U}^T \, q_c \; , \hbo
\bar{q}_c^\prime \; = \; \bar{q}_c \, {\cal U}^T \;
\hbo {\cal U} \in SU_A(2)
\label{eq:c9}
\ena
Using (\ref{eq:c2}) and (\ref{eq:c3}), one readily shows that the
kinetic energy for the charge conjugate quark fields, i.e.
$ \bar{q}_c \, \widetilde{S}_0^{-1} \, q_c $, is chiral invariant.

\vskip 0.3cm
\subsection{ Chiral behavior of diquark fields }
\label{app:c2}

In this subsection, we study the behavior of the composite field
$\bar{q}_c \; \epsilon \, \Gamma \; q $ under chiral
transformations (\ref{eq:c2},\ref{eq:c3}) and
(\ref{eq:c8},\ref{eq:c9}), respectively, where $\Gamma $ is a
Dirac matrix which will be specified when needed, and $\epsilon
_{ik}$ is the totally anti-symmetric tensor acting in flavor space
only. Note that here $q,\bar{q},q_c,\bar{q}_c$ are treated as
independent fields with the transformation properties as given
above, and that one must not use the classical equation of motion,
which e.g. yields the relation $q_c \; = \; C (q^\dagger )^T$. In
the Minkowskian formulation there would be no difference, since
chiral symmetry is manifest at classical level. As explained
above, this is not the case for the Euclidean formulation.

\vskip 0.3cm
It is convenient to introduce left- and right-handed projectors by
\be
P_L \; = \; \frac{1}{2} (1 + \gamma _5) \; , \hbo
P_R \; = \; \frac{1}{2} (1 - \gamma _5) \; , \hbo
P_L^T \; = \; P_L \; , \hbo P_R^T \; = \; P_R \; .
\label{eq:c10}
\en
The chiral matrices ${\cal V} \in SU_V(2)$ (\ref{eq:c2a}) and ${\cal U} \in
SU_A(2)$ (\ref{eq:c3a}) can be written as
\be
{\cal V} \; = \; V \times 1_D \; , \hbo
{\cal U} \; = \; U \times P_L \; + \; U^\dagger \times P_R \; ,
\label{eq:c11}
\en
where $U$ and $V$ are matrices with iso-spin indices only.

\vskip 0.3cm
Let us study chiral vector transformations first. In this case, we find
\be
\left(\bar{q}_c \; \epsilon \, \Gamma \; q \right)^\prime
\; = \; \bar{q}_c \; {\cal V}^T \, \epsilon \, \Gamma \, {\cal V} \; q
\; = \; \bar{q}_c \; \epsilon \, \Gamma \; q \; ,
\label{eq:c12}
\en
where we have used
\be
\epsilon _{ik} \, V_{il} \, V_{km} \; = \; \epsilon _{lm} \; .
\label{eq:c13}
\en
The diquark composite is invariant under chiral vector transformations.

\vskip 0.3cm
In the case of chiral axial-vector transformations, one obtains
\be
\left(\bar{q}_c \; \epsilon \, \Gamma \; q \right)^\prime
\; = \; \bar{q}_c \; {\cal U}^T \, \epsilon \, \Gamma \, U \; q \; = \;
\bar{q}_c \; \left( U^T P_L \, + \, \left(U^{\dagger }\right)^T P_R
\right) \; \epsilon \, \Gamma \, \left( U P_L \, + \, U^{\dagger } P_R
\right) \; q \; .
\label{eq:c14}
\en
In the case of an axial ($\Gamma = 1 $) or scalar
$(\Gamma = \gamma _5)$ diquark composite, we find that the axial-vector
chiral symmetry is respected, i.e.
\be
\left(\bar{q}_c \; \epsilon \, \Gamma \; q \right)^\prime
\; = \; \bar{q}_c \; U^T \; \epsilon \, U P_L  \; q \; + \;
\bar{q}_c \; \left(U^{\dagger }\right)^T \; \epsilon \,  U^{\dagger }
P_R \; q \; .
\label{eq:c15}
\en
Using (\ref{eq:c13}), the invariance under axial-vector chiral
transformations becomes obvious. In the case of an axial- vector
($\Gamma = \gamma ^\mu $), the diquark composite
is not invariant under chiral axial-vector transformations, i.e.
\be
\left(\bar{q}_c \; \epsilon \, \gamma ^\mu \; q \right)^\prime
\; = \; \bar{q}_c \; U^T \; \epsilon \, U^\dagger \gamma ^\mu  \; P_R
\; q \; + \;
\bar{q}_c \; \left(U^{\dagger }\right)^T \; \epsilon \, \gamma ^\mu  \;
U  P_L \; q \; .
\label{eq:c16}
\en
The same is true for vector diquark composites ($\Gamma = \gamma ^\mu
\gamma _5$).

\vskip 0.3cm
\setcounter{equation}{0}\renewcommand{\theequation}{\mbox{D.\arabic{equation}}}
\section{ Collective field approach to quark dynamics } 

\subsection{Fierz transformation of the current-current interaction}
\label{app:d}

As emphasized in numerous places in the main text, the dynamics in the 
$V_0$ channel associated with the ISB effect is not dominantly controlled
by the exchange of one dressed 
gluon. However in the literature, analysis of the QCD phase structure involving
color superconductivity are made with one-gluon exchange four-Fermi 
interactions.
In this appendix, we write down the various terms that result when the
current-current interaction coming from one-gluon exchange is Fierzed.
In doing this,  we shall make use of the relations 
\bea
T^A_{\alpha \beta } \, T^A_{\gamma \delta } &=& \frac{1}{2} \left(
\delta _{\beta \gamma } \, \delta _{\alpha \delta } \; - \;
\frac{1}{N} \, \delta _{\alpha \beta } \, \delta _{\gamma \delta }
\right) \; , 
\label{eq:d1} \\ 
\delta _{ik} \, \delta _{lm} &=& \frac{1}{N_f} \,
\delta _{lk} \, \delta _{im} \; + \; \frac{1}{N_f} \,
\tau ^a _{lk} \, \tau ^a_{im} \; ,
\label{eq:d2} 
\ena
where $T^A$ are $SU(N=3)$ color matrices with the proper
normalization, i.e. \break $ \tr \{ T^A T^B \} = \delta ^{AB} /2
$, and $\tau ^a$ are the $SU(N_f)$ matrices that act in flavorspace. 
These matrices are normalized according $ \tr \{ \tau ^a \tau ^b \} = 
N_f \, \delta ^{ab} $ and coincide with the Pauli matrices in the 
case of a SU(2) flavor group. 
The effective quark interaction in the soft momentum limit
gives rise to the contact interaction (\ref{eq:2.21}). Let us
concentrate on the first term of the right-hand side, i.e.
\be
\bar{q}^\alpha _i \;  \delta _{ik} \, T^a_{\alpha \beta } \,
\gamma ^\mu \; q^\beta _k \; \; \;
\bar{q}^\gamma _l \;  \delta _{lm} \, T^a_{\gamma \delta } \,
\gamma ^\mu \; q^\delta _m \; .
\label{eq:d3}
\en
Lorentz indices are not explicitly presented. Replacing $T^A
\times T^A$ in (\ref{eq:d3}) with the help of (\ref{eq:d1}), the
second expression on the right hand-side of (\ref{eq:d1}) gives
rise to the quark interaction in the vector channel, i.e.
\be
- \frac{1}{2N} \; \bar{q} \, 1_F 1_c \gamma ^\mu \, q \;
\bar{q} \, 1_F 1_c \, \gamma ^\mu q \; .
\label{eq:d4}
\en
Rearranging the product of the $\gamma $ matrices, i.e.
\bea
\gamma ^\mu _{ik} \, \gamma ^\mu _{lm} &=& - \, 1_{lk} \, 1_{im} \; + \;
\left(\gamma _5\right)_{lk} \, \left(\gamma _5\right) _{im} \;
\label{eq:d5} \\ 
&-& \frac{1}{2} \, \gamma ^\mu _{lk} \, \gamma^\mu _{im} \; - \;
\frac{1}{2} \, \left( \gamma ^\mu \gamma _5\right)_{lk} \, 
\left(\gamma ^\mu \gamma _5\right) _{im} \; , 
\nonumber 
\ena 
and using (\ref{eq:d2}), 
the first expression on the right hand side of (\ref{eq:d1}) gives
rise to an additional interaction in the vector channel. 
Taking into account a factor $(-1)$ which arises from the anti-commuting
Grassmann fields $q$ and $\bar{q}$, the vector interaction is given 
by 
\be 
\left[ \frac{1}{2} \, \frac{1}{N_f} \, \frac{1}{2} \; - \; 
\frac{1}{2N} \right] \; \bar{q} \, 1_F 1_c \gamma ^\mu \, q \;
\bar{q} \, 1_F 1_c \, \gamma ^\mu q \; .
\label{eq:d6}
\en 
One easily verifies that (\ref{eq:d6}) also holds for the one flavor 
case $N_f=1$. We observe that the 
interaction in this channel changes sign if the number of flavors is 
increased from one to two: It is repulsive for one flavor but
becomes attractive for two or more flavors if one takes three
colors. It is interesting to note that it is repulsive for any
number of flavors in the large $N$ limit. Since the large $N$ limit
is believed to give a qualitatively correct description of hadron 
interactions, it is in this limit that the {\it known} repulsion
between two nucleons in the isoscalar vector channel (i.e., 
the $\omega$ exchange channel) could be understood from the point of
view of QCD variables. In our work we are exploiting the fact that the
gluonic four-quark interaction is attractive for $N_f=2$ and $N=3$. 
But as stressed
this is not the whole story because of the possible contribution coming from
the collective degrees of freedom as discussed in \cite{KRBR}. Even if the
gluonic four-quark interaction were repulsive to start with, as long as
it is not too repulsive, it could be overpowered at some density by the 
attraction due to the ``dropping"
$V_0$ mode as the chiral phase transition is approached. This would be the
scenario if the chiral transition mechanism in terms of hadronic
variables discussed in \cite{KRBR} were realized.

\vskip 0.3cm 
The first expression on the right hand side of (\ref{eq:d1}) also 
induces interactions in the scalar and the pionic channel,
respectively, i.e.  
\be
\frac{1}{2 N_f} \, \bar{q} \, 1_F 1_c \, q
\; \bar{q} \, 1_F 1_c \, q \; - \; \frac{1}{2N_f} \, 
\bar{q} \, 1_c\gamma _5 \tau ^a \, q
\; \bar{q} \, 1_c \gamma _5 \tau ^a \, q \; .
\label{eq:d7}
\en
Note that this sign of the scalar interaction is crucial for quark 
condensation.

\vskip 0.3cm 
In the following, we specialize to the two flavor case, i.e. 
$N_f=2$. The gluon induced interaction that involves charge
conjugated quark fields is
\be
\left(\bar{q}_c\right)^\alpha _r \; \delta _{rl} \, \left(
T^A_{\alpha \beta } \right)^T
\, \gamma ^\mu \; \left(q_c\right)^\beta _l \; \;
\bar{q}_i^\gamma \; \delta _{ik} \, T^A_{\gamma \delta } \,
\gamma ^\mu \; q^\delta _k \; .
\label{eq:d10}
\en
We wish to estimate the strength this contact interaction assigns
to the quark interaction (\ref{eq:2.7}) in the (axial) vector
diquark channel. For this purpose, first note that \bea
\left(T^A\right)^T _{\alpha \beta } \, T^A_{\gamma \delta } &=&
T^A _{\beta \alpha } \, T^A_{\gamma \delta } \; = \; \frac{1}{2N}
\; \epsilon _{A\, \alpha \delta } \, \epsilon _{A \, \gamma \beta
} \; + \; \ldots \; , \label{eq:d11} \\ \gamma ^\mu _{rl} \,
\gamma ^\mu _{ik} &=& - \frac{1}{2} \; \gamma ^\nu _{il} \, \gamma
^\nu _{rk} \; - \; \frac{1}{2} \; \left( \gamma ^\nu \gamma _5
\right) _{il} \, \left( \gamma ^\nu \gamma _5 \right) _{rk}
\label{eq:d12} \\ &-& 1_{il} \, 1 _{rk} \; + \; \left(\gamma
_5\right)_{il} \, \left(\gamma _5 \right) _{rk} \; + \; \ldots \;
, \ena where we used (\ref{eq:d1}) to derive (\ref{eq:d11}). The
(axial) vector diquark composite field which we investigated in
section \ref{sec:3} is completely anti-symmetric in flavor space.
Noting that $\epsilon _{ik} = - i \tau ^{a=2} _{ik} $ and using
(\ref{eq:d2}), one finds
\be
\delta _{rl} \, \delta _{ik} \; = \; - \, \frac{1}{2} \; \epsilon _{il}
\, \epsilon _{rk} \; + \; \ldots \; .
\label{eq:d13}
\en
Combining (\ref{eq:d11}, \ref{eq:d12}, \ref{eq:d13}) and taking into
account the additional factor $(-1)$ from the interchange of
Grassmann fields, we finally obtain the desired result
\bea
&-& \frac{1}{8N}
\left(\bar{q}_c\right)^\alpha _i \, \epsilon _{ik} \epsilon ^{\gamma
\alpha \beta } \gamma ^\mu \, \left( \gamma _5 \right) \; q^\beta _k \;
\bar{q} ^\kappa _l \, \epsilon _{lm} \epsilon ^{\gamma
\kappa \omega } \gamma ^\mu \, \left( \gamma _5 \right) \;
\left(q_c\right)^\omega _m
\label{eq:d14} \\
&-&
\frac{1}{4N}
\left(\bar{q}_c\right)^\alpha _i \, \epsilon _{ik} \epsilon ^{\gamma
\alpha \beta } \; q^\beta _k \;
\bar{q} ^\kappa _l \, \epsilon _{lm} \epsilon ^{\gamma
\kappa \omega } \; \left(q_c\right)^\omega _m
\nonumber \\
&+&
\frac{1}{4N}
\left(\bar{q}_c\right)^\alpha _i \, \epsilon _{ik} \epsilon ^{\gamma
\alpha \beta }  \, \gamma _5  \; q^\beta _k \;
\bar{q} ^\kappa _l \, \epsilon _{lm} \epsilon ^{\gamma
\kappa \omega } \, \gamma _5  \;
\left(q_c\right)^\omega _m  \; .
\nonumber
\ena

\subsection{The Fierz rearrangements and approximations}

There are a variety of Fierz rearrangements that can be made on
eq.(\ref{eq:d3}) that could lead to different results depending on
approximations made in the calculations. The expressions given above
(eqs. (\ref{eq:d6}) and (\ref{eq:d7})) represent one such 
rearrangement and are exact as they stand. Note that they involve
no color generators and hence can in certain approximations be completely
saturated by color singlet meson fields. An alternative rearrangement
was used by Vogel et al \cite{vog91} wherein even the second term of 
(\ref{eq:d1}) is ``Fierzed," thereby obtaining an expression that
contains terms involving the color generators. This is also exact. Thus
{\it in the operator form}, it is identical to our Fierzed form. However
the complete saturation of this form in the same approximation as
with our expressions must require {\it colored} states in
contrast to our formulas. 

\vskip 0.3cm
An exact field theoretic treatment of the four-Fermi interaction (\ref{eq:d3})
should of course yield the same result whether one uses our expressions or
those of ref.\cite{vog91}. However if one does approximations in treating 
this {\it effective} four-Fermi interaction, there is no reason why one should
expect the same result. Indeed if one does the conventional mean-field
approximation on the two forms, one sees immediately that the results would be
different, even qualitatively in some cases. Thus a comparison of
the operators of the same form (e.g., same quantum numbers) would be
completely meaningless even though in perturbation theory, one should
get the same result (e.g., the quark propagator) order by order. 
In the next subsection, we describe the strategy -- or approximations --
that we adopt in handling the effective four-Fermi interaction that arises
when gluon-mediated interactions flow to the Fermi surface. We admit that
we have no rigorous proof that our approximation is the best for the problem
but we believe that it catches the essence of the physics we are developing
in this paper.


\subsection{ The approximation of weakly coupled mesons } 
\label{app:d2} 

\begin{figure}[t]
\parbox{6cm}{ 
\hspace{1cm} 
\centerline{ 
\epsfxsize=8cm
\epsffile{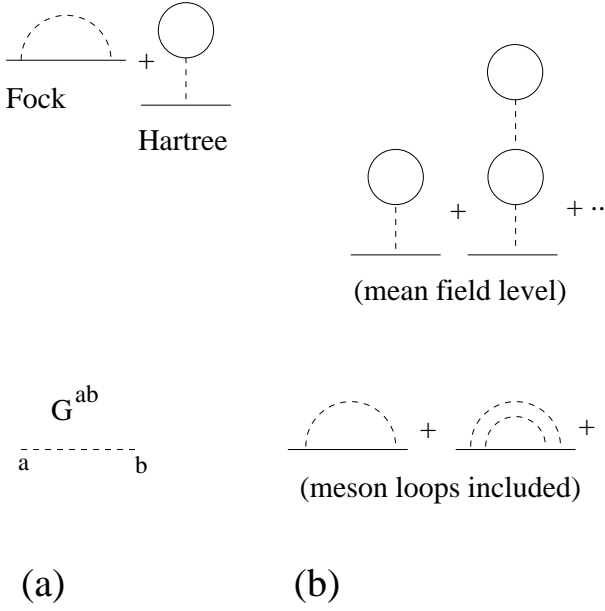} 
}
} \hspace{1.5cm}
\parbox{6cm}{ 
\caption{ Panel a: The leading order perturbative diagrams contributing 
  to the quark propagator (solid line) in bosonized quark theory; 
Panel b: leading and 
  next-to-leading order of the semi-classical expansion of bosonized 
  quark theory. 
   }\label{fig:d}

} 
\end{figure} 
Since the four-Fermi interaction (\ref{eq:d1}) 
is an effective local Lagrangian,
we would like to work with physical meson fields and approach the 
phase transition in terms of hadronic variables. For this, we find 
it convenient to set up an approximation scheme by introducing 
auxiliary fields representing quark anti-quark collective fields 
(bosonization). In order to make our argument as general as possible,
consider the general case of the nonlocal form 
\be 
S_I \; := \; \bar{q} _i \; \Gamma ^A _{ik} q_k \; G^{AB} \; 
\bar{q} _l \; \Gamma ^B _{lm} q_m \; , 
\label{eq:d2.1} 
\en 
where $k$ is a master index representing space-time as well as internal 
degrees of freedom. Calculating e.g. the quark self energy in 
perturbation theory with respect to $G^{AB}$, the $s$- and $t$-channel 
contractions contribute with equal weight and give rise (at leading order) 
to the Hartree and Fock contribution as depicted in figure \ref{fig:d}a. 

\vskip 0.3cm 
Now such nonperturbative phenomena as quark 
condensation or dynamical mass generation take place 
in a strong-coupling regime which cannot be accessed by perturbation theory. 
To describe such phenomena, it is more advantageous to
rewrite the interaction (\ref{eq:d2.1}) 
by means of auxiliary meson fields, i.e. 
\bea
\exp \{ - S_I \} &=& \int {\cal D} \Sigma \; \exp \{ - S_M \} \; , 
\label{eq:d2.2} \\ 
S_M &=&  
\frac{1}{2} \Sigma _{ik} \, \left( M^{-1}\right)_{ik,lm} \, \Sigma _{lm} 
\; + \; i \, \bar{q}_l q_m \, \Sigma _{lm} \; , 
\label{eq:d2.3} \\ 
M_{ik,lm} &=& 2 \, \Gamma ^A _{ik} \, G^{AB} \, \Gamma ^B_{lm} \; . 
\label{eq:d2.4} . 
\ena 
The complete partition function can then be written in the form 
\be 
Z \; = \; \int {\cal D} q \; {\cal D} \bar{q} \; {\cal D} \Sigma \; 
\exp \left\{ - \bar{q}_i \; \left(S^{-1}_0\right)_{ik} q_k \; - S_M \, 
\right\} \; , 
\label{eq:d2.5} 
\en 
where $S^{-1}_0 = i \dslash -m $ is the tree-level inverse quark propagator. 
Non-perturbative effects can be addressed, by integrating out the quark 
fields and treating the  functional integral over the ``meson" field 
$\Sigma $ in a semi-classical 
approximation. Expanding the action of the effective meson theory up to 
second order in the meson field $\Sigma = \sigma + \Sigma _0$ yields 
\bea 
Z &=& \int {\cal D} \sigma \; \exp \left\{ - S_{eff} \right\} \; , 
\label{eq:d2.6} \\ 
S_{eff} &=& 
\frac{1}{2} \sigma _{ik} \, \left( M^{-1} \, + \, \Pi \right)_{ik,lm} 
\, \sigma _{lm} \; - \; S_{ml} \, \sigma _{lm} 
\; - \; \ln S^{-1} \; + \; {\cal O}\left(\sigma ^3 \right) \; , 
\label{eq:d2.7} 
\ena 
where $S_{ml} $ is the ``constituent" quark propagator with $\Sigma _0 $ 
as self-energy, i.e. 
\be 
S^{-1} \; = \; i \dslash \, - \, m \, + \, i \Sigma _0 \; , 
\label{eq:d2.8} 
\en 
and 
\be 
\Pi _{ik,lm} \; = \; S_{kl} \, S_{mi} \; . 
\label{eq:d2.9} 
\en 
If one chooses in the bosonized action 
$\Sigma _0$ in such a way that the linear terms in 
$\sigma $ vanish, the quark self-energy will be given by 
a ladder summation of the Hartree terms as shown in the upper
graph of figure \ref{fig:d}b. 
The Fock terms of the perturbative expansion are generated, if 
Gaussian fluctuations $\sigma $ around the mean field $\Sigma _0$ 
are included as in the lower graph of figure \ref{fig:d}b. 

\vskip 0.3cm 
There are several options available depending on how one bosonizes
the interaction (\ref{eq:d3}).
\begin{enumerate}
\item  One may choose to bosonize eq.(\ref{eq:d3}) directly and then
do a semi-classical approximation in the mesonic functional 
integral. This procedure would however be a poor approximation for the physics
we are interested in.
The reason is that due to vanishing Hartree terms in this case, at
the classical level, it would yield a trivial result for the quark 
self-energy, whereas the nontrivial phenomena (such as quark
condensation, constituent quark mass 
generation etc.) would occur at semi-classical level or higher. 
\item
Suppose one Fierz transforms (partially), thereby rearranging the interaction 
terms (partially). In this case,  the role of the 
Fock and Hartree terms involved will be (partially) interchanged. 
Subsequent bosonization and semi-classical 
approximation of the mesonic functional integral would correspond 
to a particular summation of selected perturbative diagrams. Now the 
crucial point is that  
different choices of the Fierz rearrangement would differ in the subset 
of Fock and Hartree diagrams which are summed up. Such different choices
would correspond to doing different physics. The examples are the choice
adopted in this paper and that adopted in \cite{vog91}.
\begin{itemize}
\item
The first case to
consider is the {\it complete} Fierz rearrangement of the interaction 
(\ref{eq:d3}) studied in \cite{vog91}. This choice would reproduce 
the important Fock terms at tree level. Doing (weak coupling) perturbation 
theory would bring minor corrections from the meson fluctuations. 
Although the semi-classical expansion 
induced by this choice of Fierz rearrangement may be rapidly 
converging in weak-coupling regime, this must not be the 
case for the coupling strength where quark condensation effects 
are expected to come into play. Indeed, an evidence for this shortcoming
can be found in~\cite{ebe86}: the Fierz 
rearrangement of~\cite{vog91} implies a ratio of the interaction strengths 
of scalar and vector channel, respectively, of $G_s/G_v = 2$ whereas 
a ratio of $G_s/G_v \ll 1$ seems to be needed to reproduce the 
$\pi $ and $\rho $ meson masses~\cite{ebe86}. 
\item
Consider next the Fierz rearrangement adopted in this paper. We argue that
doing mean-field approximation with this form corresponds to doing
the large $N_c$, $N_f$ approximation, i.e. 
\be 
N_f \; = \; \nu \, N_c \; , \hbo \nu \, = \, \hbox{fixed const} \, , \; 
N_c \rightarrow \infty \;.  
\label{eq:d2.10} 
\en 
In this limit, the effect of meson fluctuations will be suppressed\footnote{
To those who are familiar with Walecka mean field theory of nuclear matter,
this procedure is easily comprehensible. One could interpret this in terms
of Landau's Fermi-liquid fixed point theory with the instability associated
with the ISB and color superconductivity interpreted as the break-down of
the Fermi-liquid ground state. See \cite{friman} for a discussion on 
this matter.}. 
Note that the limit 
(\ref{eq:d2.10}) does not spoil the property of asymptotic freedom 
of Yang-Mills theory for $\nu < 11/2$ (note that one has $\nu =1$ 
for QCD with three quark flavors). In the particular case (\ref{eq:d2.10}), 
the four quark interaction strength is solely proportional 
to $1/N_c$. 
Bosonization of these interaction terms gives rise to large meson masses, 
i.e. proportional to $N_c$. Mesonic kinetic terms are produced by the 
logarithm of the quark determinant. These kinetic terms therefore also 
acquire a large pre-factor of order $N_c$. These considerations tell us 
that, in our case of the Fierz rearrangement, meson fluctuations are 
suppressed by a factor of $1/N_c$ 
thereby justifying our semi-classical expansion in the particular limit 
(\ref{eq:d2.10}). 

\vskip 0.3cm 
For practical applications ($N_c=2$), 
it is not clear whether a truncation of the 
large $N_c$, $N_f$ expansion (\ref{eq:d2.10}) at classical level is a
truly good approximation. 
In particular at zero density, a higher-order approximation
may be needed to 
generate the masses for the vector mesons. Rather than justifying 
the large $N_c$, $N_f$ expansion for the case of interest, e.g.
$N_c=3$, $N_f=2$, we here adopt a more coarse-grained point of view 
and argue that in view of the results in~\cite{KRBR} (see discussion 
in section \ref{sec:dis}) it seems worthwhile from a phenomenological 
point of view to introduce in the case of finite densities the mean-field 
$V_0$ and leave the strength of the $V_0$ fluctuations as a variable
parameter. 
The goal of our paper is to explore the behavior of the finite density 
quark matter by varying the strength in this particular channel 
rather than relying on the strength provided by 
the effective four-Fermi interaction of one-gluon exchange 
quantum numbers (\ref{eq:d3}). 
\end{itemize}\end{enumerate}

\vskip 0.3cm
\setcounter{equation}{0}\renewcommand{\theequation}{\mbox{E.\arabic{equation}}}
\section{ Fermion determinants }
\label{app:e}

Let us first quote two useful results, which will be used below, i.e.
\bea
\dete_D \left( \alpha \kslash + \beta \right) &=&
\left[ \alpha^2 k^2 + \beta ^2 \right]^2 \; ,
\label{eq:e1} \\
\dete_D \left( \epsilon \gamma ^0 \kslash ^+ \gamma ^0 \, + \, \alpha
\kslash ^- \, + \, \beta \right) &=&
\left[ (\epsilon k_0^+ - \alpha k_0^-)^2 \, + \, (\epsilon + \alpha )^2
\vec{k}^2 \, + \, \beta ^2 \, \right]^2 \; ,
\label{eq:e2}
\ena
where $\alpha $, $\beta $, $\epsilon $ are constants, and
$k^\pm := ( k_0^\pm , \vec{k} )^T$, $k_0^\pm = k_0 \pm i v $.
The subscript $D$ indicates that the determinant is over Dirac indices.
The equations (\ref{eq:e1},\ref{eq:e2}) can be readily checked e.g.
by resorting to the explicit representation of the $\gamma $--matrices
(\ref{eq:b4}).

\vskip 0.3cm
The determinants of interest possess the structure
\be
T:\; = \; \dete _{cDCF} \left(
\matrix{ (\kslash ^+ - \sigma ) \, 1_C \, 1_F & \kappa \Delta ^\dagger \cr
\kappa \Delta & (\kslash ^- -\sigma )  \, 1_C \, 1_F }
\right) \; ,
\label{eq:e3}
\en
where the building blocks of the $2\times 2$ matrix in the $q$ and $q_c$
space carry Dirac, flavor and color indices. The subscript $cDCF$ indicates
the determinant extends over either space, i.e.
$c$ charge conjugated space, $D$ space of Dirac indices, $C$
color and $F$ flavor space.
Exploiting the simple structure of the diagonal entries, $T$ can be
reduced to
\bea
T &=& \left(k_+^2 +\sigma ^2 \right)^{12} \;
\dete _{DCF} \left[ \kslash ^- \, - \, \sigma \, - \, \kappa ^2 \,
\Delta \, \left( \kslash ^+ - \sigma \right) ^{-1} \, \Delta ^\dagger
\right]
\nonumber \\
&=& \left(k_+^2 +\sigma ^2 \right)^{12} \;
\dete _{DCF} \left[ \kslash ^- \, - \, \sigma \, + \, \frac{
\kappa ^2 \, \delta ^2 }{
k_+^2 + \sigma ^2 } \, \Gamma \, \left( \kslash ^+ + \sigma \right) \,
\Gamma ^\dagger \right] \,
\label{eq:e4}
\ena
where we have used $\Delta = \delta \, \Gamma $. Let us specify the
color, Dirac and flavor structure of the diquark vertex, i.e.
\be
\Gamma ^{\alpha \beta } _{ik} \; = \; \epsilon ^{\alpha \beta 3 }
\, \epsilon _{ik} \; \gamma _D \; ,
\label{eq:e5}
\en
where $\alpha , \beta =1 \ldots 3$ are color indices, and $i,k=1,2$
are flavor indices, respectively. Dirac indices are not explicitly
shown. The Dirac structure will be specified below.
Since the propagator $( \kslash ^+ - \sigma ) ^{-1}$ is diagonal
in color and flavor space, one obtains
\be
\Gamma \left( \kslash ^+ - \sigma \right) ^{-1} \Gamma ^\dagger \; = \;
\gamma _D \left( \kslash ^+ - \sigma \right) ^{-1} \gamma _D^\dagger
\, \times \, 1_F \, \times \, P_C \; ,
\label{eq:e6}
\en
where $1_F$ is the unit matrix in $2\times 2$ flavor space, and
$P_C= \hbox{diag}( 1,1,0)$ is a projector in color space.
We therefore obtain
\bea
T &=& \left(k_+^2 +\sigma ^2 \right)^{12} \;
\left(k_-^2 +\sigma ^2 \right)^{4} \;
\label{eq:e7} \\
&& \left\{
\dete _{D} \left[ \kslash ^- \, - \, \sigma \, + \, \frac{
\kappa ^2 \, \delta ^2 }{
k_+^2 + \sigma ^2 } \, \gamma _D \, \left( \kslash ^+ + \sigma \right)
\, \gamma _D ^\dagger \right] \, \right\} ^4 \; .
\nonumber
\ena
Investigating the various types diquark composites, i.e.

\vskip 0.3cm
\begin{tabular}{ll}
\hspace{3cm} $ \gamma_D  \; = \; \gamma _5 $ \hspace{3cm} & (scalar) \cr
\hspace{3cm} $ \gamma _D\; = \; 1_D \, $ \hspace{3cm} & (pseudo-scalar) \cr
\hspace{3cm} $ \gamma _D \; = \; \gamma ^0 \gamma _5 $ \hspace{3cm} &
(vector) \cr
\hspace{3cm} $ \gamma _D \; = \; \gamma ^0 $ \hspace{3cm} & (axial-vector),
\end{tabular}

\vskip 0.3cm
it is straightforward with the help of (\ref{eq:e1},\ref{eq:e2})
to calculate the final result for the fermion determinants
(\ref{eq:3.3}), (\ref{eq:3.10}), (\ref{eq:3.22}), (\ref{eq:3.22b}).

\end{document}